\documentclass[twocolumn]{aastex63}
\usepackage{lineno, blindtext}
\usepackage{epsfig}
\usepackage{empheq}
\usepackage{natbib}
\usepackage{amsmath}
\usepackage{verbatim}
\newcommand{\DIW}{$\Delta$IW }
\usepackage{flafter}
\usepackage{float}
\usepackage{graphicx}
\usepackage{amssymb}
\usepackage{wasysym}
\usepackage{xfrac}
\usepackage{footmisc}
\usepackage{booktabs}
\makeatletter
\def\blfootnote{\xdef\@thefnmark{$^\ast$}\@footnotetext}
\makeatother

\shorttitle{New chondritic bodies in oxygen-bearing white dwarfs}
\shortauthors{A. E. Doyle et al.}

\begin{document}

\title{New chondritic bodies identified in eight oxygen-bearing white dwarfs}

\correspondingauthor{Alexandra E. Doyle, Edward D. Young}
\email{a.doyle@epss.ucla.edu, eyoung@epss.ucla.edu}

\author[0000-0003-0053-3854]{Alexandra E. Doyle}
\affiliation{Earth, Planetary, and Space Sciences, University of California, Los Angeles, Los Angeles, CA 90095, USA}

\author[0000-0001-5854-675X]{Beth L. Klein}
\affiliation{Department of Physics and Astronomy, University of California, Los Angeles, CA 90095-1562, USA}

\author[0000-0003-4609-4500]{Patrick Dufour}
\affiliation{Institut de Recherche sur les Exoplan\`etes (iREx), Universit\'e de Montr\'eal, Montr\'eal, QC H3C 3J7, Canada}
\affiliation{D\'epartement de Physique, Universit\'e de Montr\'eal, Montr\'eal, QC H3C 3J7, Canada}

\author[0000-0001-9834-7579]{Carl Melis}
\affiliation{Center for Astrophysics and Space Sciences, University of California, San Diego, CA 92093-0424, USA}

\author[0000-0001-6809-3045]{B. Zuckerman}
\affiliation{Department of Physics and Astronomy, University of California, Los Angeles, CA 90095-1562, USA}

\author[0000-0002-8808-4282]{Siyi Xu}
\affiliation{Gemini Observatory/NSF’s NOIR Lab, Hilo, HI, 96720, USA}

\author[0000-0001-6654-7859]{Alycia J. Weinberger}
\affiliation{Earth and Planets Laboratory, Carnegie Institution for Science, 5241 Broad Branch Rd NW, Washington, DC 20015, USA}

\author[0000-0002-1299-0801]{Isabella L. Trierweiler}
\affiliation{Earth, Planetary, and Space Sciences, University of California, Los Angeles, Los Angeles, CA 90095, USA}
\affiliation{Department of Physics and Astronomy, University of California, Los Angeles, CA 90095-1562, USA}

\author{Nathaniel N. Monson}
\affiliation{Earth, Planetary, and Space Sciences, University of California, Los Angeles, Los Angeles, CA 90095, USA}

\author{Michael A. Jura$^\ast$}
\affiliation{Department of Physics and Astronomy, University of California, Los Angeles, CA 90095-1562, USA}
\blfootnote{Deceased}

\author[0000-0002-1299-0801]{Edward D. Young}
\affiliation{Earth, Planetary, and Space Sciences, University of California, Los Angeles, Los Angeles, CA 90095, USA}
 
\begin{abstract}
We present observations and analyses of eight white dwarf stars that have accreted rocky material from their surrounding planetary systems. The spectra of these helium-atmosphere white dwarfs contain detectable optical lines of all four major rock-forming elements (O, Mg, Si, Fe). This work increases the sample of oxygen-bearing white dwarfs with parent body composition analyses by roughly thirty-three percent. To first order, the parent bodies that have been accreted by the eight white dwarfs are similar to those of chondritic meteorites in relative elemental abundances and oxidation states. Seventy-five percent of the white dwarfs in this study have observed oxygen excesses implying volatiles in the parent bodies with abundances similar to those of chondritic meteorites. Three white dwarfs have oxidation states that imply more reduced material than found in CI chondrites, indicating the possible detection of Mercury-like parent bodies, but are less constrained. These results contribute to the recurring conclusion that extrasolar rocky bodies closely resemble those in our solar system, and do not, as a whole, yield unusual or unique compositions.
\end{abstract}

\section{Introduction \label{intro}}

Categorization of the compositions of rocky exoplanets, and evaluation of their similarities to or differences from rocky bodies in our solar system, is a challenging and flourishing area of study. To this end, many studies have characterized exoplanet compositions using stellar spectroscopy of FGK, or Sun-like, stars \citep[e.g.,][]{unterborn2019,adibekyan2021,kolecki2022} in combination with planetary mass-radius relations. An alternative approach is to use white dwarf stars (WDs) -- stars in the last stage of stellar evolution -- that have been ``externally-polluted" by accretion of rocky bodies from their surrounding planetary systems. Owing to their strong gravitational acceleration, the atmospheres of WDs are typically devoid of elements heavier than helium. The heavy elements sink out of the observable atmosphere on timescales of days to millions of years \citep{Koester2009}, depending on the atmospheric temperature and dominant constituent (H or He). Because of the relatively short settling timescales of heavy elements, externally-polluted WDs must have acquired their heavy elements relatively recently compared to their lifetimes.  Radiative levitation as a mechanism to maintain heavy elements in a white dwarf atmosphere \citep[e.g.,][]{chayer1995} is not effective for the white dwarfs presented herein (helium atmosphere WDs with effective temperatures cooler than 20,000K).

WDs for which hydrogen presents the strongest spectral line are referred to as `DAs' and neutral helium as `DBs.' If a spectrum displays both H {\small I} and He {\small I} lines, the spectral type can be either DAB or DBA depending on whether H or He, respectively, has the strongest optical absorption line. White dwarfs are deemed polluted if any element heavier than He is detected in their atmosphere; following \citet{sion1983} and \citet{wesemael1993}, we denote external-pollution with a ‘Z’ in the spectral classifications.

We now understand that these polluted WDs, constituting 25-50\% of all WDs, accrete material from the planets, asteroids, and comets that orbited the host star and were subsequently scattered toward the star by the post-main sequence evolution \citep{debsig2002,RN7, RN16, RN17, RN19, veras2016}. Observations of transiting debris from planetary material that has been tidally disrupted by the WD \citep{Vanderburg2015,RN15,Vanderbosch2019,guidry2021,vanderbosch2021} suggest the presence of a body in the process of being pulverized and accreted by the WD, thus substantiating our understanding of the source of pollution. Analyses of polluted WDs to evaluate the compositions of extrasolar rocky bodies have proliferated in the last decade \citep[e.g.,][]{RN20,klein2010,Vennes2010,Melis2011,Farihi2011,Zuckerman2011,RN8,RN1,gaensicke2012cos,jurayoung2014,xu2017,RN51,Hollands_2018,Doyle_2019,Swan_2019,Bonsor2020,buchan2022}.

To date, the parent bodies being accreted by polluted WDs mostly resemble dry, rocky bodies similar in size and general composition to asteroids in the solar system. However, a few water-rich bodies \citep{Farihi2011,farihi2013,raddi2015,hoskin2020,klein2021}, including a Kuiper Belt analog \citep{xu2017}, have been discovered. Additionally, parent bodies that resemble giant planets \citep{gansicke2019} and icy moons \citep{Doyle2021} have been argued. While just a few dozen WDs are `heavily' polluted, with more than a few rock-forming elements detected, taken together, 23 distinct elements have been detected in polluted WDs \citep[see Table 1 of][]{klein2021}. Compositional variations due to igneous differentiation-- with compositions that range from crust-like to core-like -- have been identified \citep[e.g.][]{Zuckerman2011,Melis2011,gaensicke2012cos,jurayoung2014,melisdufour2017, putirka2021,hollands2021,johnson2022}.

In this work we present new observations of eight heavily polluted DB WDs and examine the compositions of the accreting rocky parent bodies. We focus on evaluating these bodies through bulk composition and oxidation state. In addition to Ca and the four major rock-forming elements (O, Mg, Si, and Fe), instances of additional elements (e.g., Al, Cr, and Ti) have been detected in some of the WDs. These new data increase the sample of oxygen-bearing WDs with parent body composition analyses by $\thicksim$ 33\%. This paper is organized as follows: in Section \ref{sec:observations} we list our target selection and observations for the WDs described. Our atmosphere models are described in Section \ref{DataAnalysis} along with spectra of the detected major rock-forming elements. Section \ref{discussion} provides an analysis of the  parent body compositions and in Section \ref{conclusions} we summarize our findings. 

\begin{figure}
\includegraphics[width=85mm]{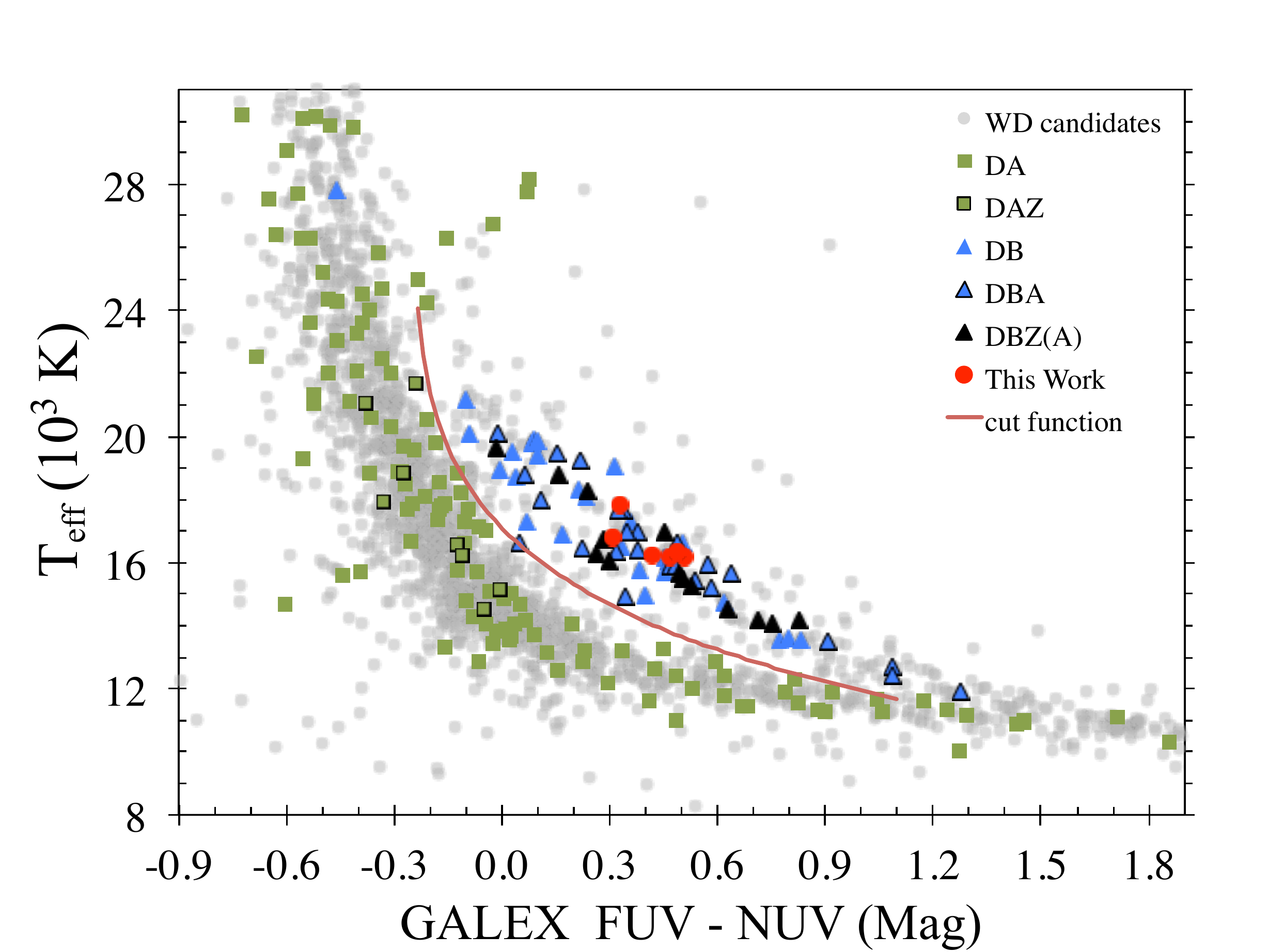}
\caption{$T_{\rm eff}$ as a function of GALEX colors. Here we show the Gaia WD candidates from \citet{gentilefusillo2019} that have both far-UV (FUV) and near-UV (NUV) GALEX data, which reveals a distinct dichotomy between DA and DB WDs. The subset of polluted helium-dominated atmospheres from this work is represented as red circles. The red curve is our constructed ``cut function," which we use to assign likely dominant elements based on the location of the WD parameters on this figure (see also Equation \ref{eqn:flags}).   
}
\label{fig:GALEX_compare}
\end{figure}

\section{Observations \label{sec:observations}}
\subsection{Target Selection \label{targets}}
In this paper we focus on eight DB WDs (Table \ref{observations}). In each of these WDs, all four major rock-forming elements (O, Mg, Si, Fe) are detected. 

Three out of eight WDs in this work have been observed over the years by members of our team. In particular we obtained HIRES spectra of WD1244$+$498 and SDSSJ1248$+$1005 because they were previously identified as DBZs in SDSS spectra \citep{kleinman2013, koesterkepler2015}, and WD1415$+$234 was followed up at high resolution due to the possible appearance of a Ca {\small II} K line in \citet{limoges2010}.

The other five WDs were identified in a search for heavily polluted WDs \citep{melis2018}. We compiled our list of targets by utilizing the sample of probable WDs from \citet{gentilefusillo2019}, which calculates stellar parameters and the probability of an object being a WD based on fits to Gaia DR2 data.

\begin{table*}[h]
\caption{WD Observation Data}
\label{observations}
\begin{center}
\begin{tabular}{lccccc}
\hline
\hline
Name & UT Date  & Instrument & Coverage (\AA) & Int. Time (sec) & SNR$^a$ \\
\hline
GaiaJ0218$+$3625 & 2021/08/31 & HIRES (blue) & 3115$-$5950 & 3600 & 43 \\
 & 2020/10/08 & HIRES (blue)&4000$-$5950&2000$\times$2&7$^{bc}$\\
 & 2019/07/16 & HIRES (red) &4715$-$8995&3300&33 \\
 & 2018/12/30 & Kast &3420$-$5485, 5590$-$7840&3300&62\\
WD1244$+$498 & 2018/05/18 & HIRES (blue) &3115$-$5950&2000&24\\
 & 2015/04/09 & HIRES (red) &4715$-$8995&1800$\times$2&39$^b$\\
 & 2010/03/28 & HIRES (blue) &3115$-$5950&3000&25\\
SDSSJ1248$+$1005  & 2015/04/09 & HIRES (red) &4715$-$8995&3000$\times$2&24$^b$\\
 & 2014/05/22 & HIRES (blue) &3115$-$5950&3000$\times$3&32$^b$\\
WD1415$+$234 & 2019/07/16 & HIRES (red) &4715$-$8995&3300&43\\
 & 2016/04/01 & HIRES (blue) &3115$-$5950&2400$\times$2&40$^b$\\
 & 2015/04/25 & ESI &3900$-$10900&1180$\times$2&25$^b$\\
SDSSJ1734$+$6052 & 2019/09/07 & HIRES (red) &4715$-$8995&3600&29\\
 & 2019/07/16 & HIRES (red) &4715$-$8995&3300&34\\
 & 2019/07/07 & HIRES (blue) &3115$-$5950&3300&27\\
 & 2019/05/29 & Kast &3415$-$5480, 6420$-$8790&3900&28\\
GaiaJ1922$+$4709 & 2020/10/07 & HIRES (red) &4715$-$8995&3600&37\\
 & 2020/06/14 & HIRES (blue) &3115$-$5950&3000&28\\
 & 2019/12/09 & HIRES (red)&4715$-$8995&3600&26\\
 & 2019/10/12 & Kast &3420$-$5485, 6400$-$8800&3000&42\\
EC22211$-$2525 & 2021/08/31 & HIRES (blue) &3115-5950&3300&40\\
 & 2020/10/07 &  HIRES (red) &4715$-$8995&3600&46\\
 & 2019/07/07 & HIRES (blue) &3115$-$5950&3300&38\\
 & 2019/07/03 & MagE &3065$-$9470&1200$\times$2&78$^b$\\
 & 2018/12/12 & Kast &3450$-$5475, 5590$-$7840 &2700&17 \\
SDSSJ2248$+$2632 & 2019/09/07 & HIRES (red) &4715$-$8995&3300&38\\
 & 2019/07/16 & HIRES (red) &4715$-$8995&3300&43\\
 & 2019/07/07 & HIRES (blue) &3115$-$5950&3000&36\\
 & 2017/12/11 & Kast &3430$-$5500, 5625$-$7820&3600&62\\
\hline
\end{tabular}
\end{center}
\raggedright $^a$ Signal-to-noise-ratio (SNR) measured at 3445\AA\ for HIRES (blue), 5195\AA\ for HIRES (red), 5160\AA\ for MagE, 5100\AA\ for Kast, and 6000\AA\ for ESI
\\$^b$ SNR for combined exposures
\\$^c$ Only CCDs 2 and 3 were used in our analysis
\end{table*}

To focus on finding DB WDs, we compared GALEX colors \citep{bianchi2017} to effective temperature ($T_{\rm eff}$) (Figure \ref{fig:GALEX_compare}). Differences in opacity of DA and DB WDs have a salient effect on emergent fluxes, particularly at UV wavelengths as observed with GALEX. These colors reveal a distinct dichotomy between DA and DB WDs \citep[e.g.][]{bergeron2019}. We constrained Gaia WD candidates from \citet{gentilefusillo2019} to include only those where Gmag $<$ 17.0, distance $<$ 300 pc, and far-UV (FUV) and near-UV (NUV) GALEX data exist, (see Figure \ref{fig:GALEX_compare}). Known characterizations of each WD are labeled as either green squares (DAs) or blue triangles (DBs), and unconfirmed WD candidates are labeled as gray circles. The polluted DBs analyzed in this paper are represented as red circles. 

To process these data for our purposes, we constructed a ``cut function" (red curve in Figure \ref{fig:GALEX_compare}) with the equation

\small
\begin{equation}
    T_{\rm eff, cut} = 28000\exp\left({-\left(\frac{{\rm FUV-NUV}+0.24}{2}\right)^{1/3}}\right),
    \label{eqn:flags}
\end{equation}

\normalsize
\noindent  that applies for $12000<T_{\rm eff}<24000$.  We used equation \ref{eqn:flags}  to flag points  as ``likely DBs" where $T_{\rm eff} > T_{\rm eff, cut}$ (above the red curve in Figure \ref{eqn:flags}) and those where $T_{\rm eff} < T_{\rm eff, cut}$ as ``likely DAs" (below the red curve, Figure \ref{eqn:flags}). This allowed us to specifically target WDs that fell within the range of known DBs. This particular selection method for observing WDs led to the discovery of many of the polluted DB WDs in this study, as well as others to be published in future studies.

\subsection{Instrument Setup \label{Instruments}}
Table \ref{observations} lists our target WDs along with their observation dates, instruments, and resulting data properties. We describe each instrument and observational setup in more detail below.

\subsubsection{KAST \label{KAST}}
Our large-scale survey to search for heavily polluted WDs from Gaia DR2 WD candidates (described in Section \ref{targets} and \citealt{melis2018}) utilized the KAST Spectrograph on the 3m Shane telescope at Lick Observatory. Our standard setup implemented the d57 dichroic, which split blue light through the 600/4310 grism and red light through the 830/8460 grating. This setup provides a resolving power (R = $\lambda$/$\Delta\lambda$) for a 2\arcsec~slit in blue and red of R = 950 and 1,500, respectively, and wavelength coverage from 3450-7800\,{\AA}. Where indicated in Table \ref{observations}, we implemented another version of our setup which tilted the 830/8460 grating to cover redder wavelengths and specifically the Ca infrared triplet ($\lambda$ 8498/8542/8662\,{\AA}) resulting in red arm wavelength coverage from 6440$-$8750\,{\AA}. For both setups, we used slit widths of 1, 1.5, or 2\arcsec~and integration times from 45$-$60 minutes depending on observing conditions and target brightness. The data were reduced using standard IRAF routines, including bias subtraction, flat-fielding, wavelength calibration using arc lamps, and instrumental response calibration using observations of standard stars \citep{iraf1986}. Signal-to-noise-ratios (SNRs) for the resulting spectra are measured at 5100\AA\, and reported in Table \ref{observations}.

\subsubsection{MagE \label{MAGE}}
Moderate resolution optical spectra of EC22211$-$2525 were acquired with the Magellan Echellette (MagE) spectrograph on the Magellan 1 (Baade) telescope at Las Campanas Observatory on 2019 July 03. EC22211$-$2525 was observed through the 0.5\arcsec~slit providing a resolving power of R $\simeq$ 7,500. Data reduction was performed with the Carnegie Python pipeline \citep{kelson2000, kelson2003} and SNR measurements were made at 5160\,{\AA}.

\subsubsection{ESI \label{ESI}}
We used the Echellette Spectrograph and Imager (ESI) on the Keck II Telescope at Maunakea Observatory \citep{sheinis2002} to obtain a spectrum for WD1415$+$234. ESI data were taken with a 0.3\arcsec slit providing a resolving power of R $\simeq$ 13,000. Data were reduced using MAKEE and IRAF, similar to the HIRES reduction process described in \citet{klein2010}. SNR for the resulting combined spectrum was $\thicksim$ 25, measured at 6000\AA.

\subsubsection{HIRES \label{HIRES}}
We used HIRES on the Keck I Telescope at Maunakea Observatory \citep{vogt1994} to obtain higher resolution spectra for each of the eight WDs in this sample. HIRES data were taken with the C5 decker (slit width 1.148\arcsec) for a resolving power of R $\simeq$ 37,000 and resulting in wavelength coverage of 3115-5950\,{\AA} with the blue collimator and 4715-8995\,{\AA} with the red collimator. Exposure times ranged from 30$-$60 minutes and depended on observing conditions and target brightness. Data were reduced using either the MAKEE software package with IRAF continuum normalization or IRAF reduction routines (see \citealt{klein2010} for more details on the methods and routines used). The SNR for the resulting spectra were measured at 3445\AA\ for HIRES blue and 5195\AA\ for HIRES red, and are displayed in Table \ref{observations}.

\begin{table*}
\caption{WD Parameters \label{parameters}}
\begin{center}
\scriptsize
\begin{tabular}{lccccccccccccc}
\hline 
\hline
WD	&	RA	&	Dec	&	G	&	D	&	Teff	&	log{\it g}	&	H$\alpha$	&	H$\alpha$	&	H$\alpha$	&	CaK	&	CaK	&	CaK	&	Spectral	\\
Name	&	(J2000)	&	(J2000)	&	(mag)	&	(pc)	&	(K)	&	(cgs)	&	EW	&	depth	&	depth	&	EW	&	depth	&	depth	& Type	\\
	&		&		&		&		&		&		&	(mA)	&	HIRES	&	lowres	&	(mA)	&	HIRES	&	lowres	&		\\
\hline																											
GaiaJ0218+3625	&	02 18 16.64	&	+36 25 07.6	&	16.4	&	116	&	14700	&	7.86	&	475	&	0.10	&	0.06	&	595	&	0.65	&	0.13	&	DBZA	\\
WD1244+498	&	12 47 03.28	&	+49 34 23.5	&	16.6	&	120	&	15150	&	7.97	&	1600	&	0.26	&	0.16	&	664	&	0.67	&	0.20	&	DBAZ	\\
SDSSJ1248+1005	&	12 48 10.23	&	+10 05 41.2	&	17.4	&	164	&	15180	&	8.11	&	1750	&	0.28	&	0.17	&	1245	&	0.66	&	0.39	&	DBAZ	\\
WD1415+234	&	14 17 55.37	&	+23 11 36.7	&	16.6	&	127	&	17300	&	8.17	&	1150	&	0.23	&	0.15	&	274	&	0.63	&	0.07	&	DBAZ	\\
SDSSJ1734+6052	&	17 34 35.75	&	+60 52 03.2	&	16.9	&	150	&	16340	&	8.04	&	2000	&	0.25	&	0.21	&	256	&	0.67	&	0.08	&	DBAZ	\\
GaiaJ1922+4709	&	19 22 23.41	&	+47 09 45.4	&	16.6	&	127	&	15500	&	7.95	&	510	&	0.16	&	0.08	&	528	&	0.57	&	0.18	&	DBZA	\\
EC22211$-$2525	&	22 23 58.39	&	$-$25 10 43.6	&	16.3	&	109	&	14740	&	7.89	&	1500	&	0.24	&	0.22	&	710	&	0.68	&	0.17	&	DBAZ	\\
SDSSJ2248+2632	&	22 48 40.93	&	+26 32 51.6	&	16.4	&	123	&	17370	&	8.02	&	750	&	0.18	&	0.15	&	169	&	0.55	&	0.07	&	DBAZ	\\
\hline
\end{tabular}
\end{center}
\tablecomments{$\rm{G}_{mag}$ and distances (calculated from parallaxes) are from Gaia EDR3 \citep{gaiacollaboration2016,gaiacollaboration2021}. $\it{T}_{\rm eff}$ and log{\it g} are fit as described in Section \ref{abundmodel}.   Typical uncertainties for $\it{T}_{\rm eff}$ and log{\it g} are $\pm$500K and $\pm$0.05, respectively. ``lowres'' refers to either SDSS or Kast spectra. Line ``depth'' is the position of the line center between the continuum and zero, measured as the fractional distance below the continuum.  Spectral Type assignments are based on equivalent widths (EWs) of Ca {\small II} K (CaK) and H$\alpha$ as described in Section \ref{spectype}. } 
\end{table*}

\section{Data Analysis \label{DataAnalysis}}
\subsection{Spectral Typing \label{spectype}}
WD spectral types are established according to the appearance of their optical spectra and do not always reflect the dominant atmospheric composition \citep[e.g.~GD 16 and GD 362,][]{koester2005, RN20}.  A colleague prudently pointed out, ``Annie Jump Cannon was prophetical when she made it clear that stellar spectral types should never have physical interpretations, because she realized models would change but spectral morphology would be static for a given type'' (J. Farihi 2022, private communication).

Three stars in our sample (WD1244+498, SDSSJ1248+1005, WD1415+234) were previously known WDs; the other five are newly identified in this work. In all cases, as of the date of this publication, the spectral types in SIMBAD are either absent or need updating.

In trying to determine the appropriate spectral types for this set of WDs, we ran into a matter that requires some clarification.  In all these spectra, the He {\small I} lines are clearly the dominant optical features: He {\small I} $\lambda$5876~\AA\ equivalent widths (EWs) range from 5-14 {\AA}, and line depths (as defined in Table 2 note) range from 0.34-0.48, with little depth difference between low and high resolution spectra. Thus the primary spectral type begins with `DB' in each case (Table \ref{parameters}).  However, since each WD also displays H$\alpha$ and high-Z lines, the question is how to distinguish whether the secondary type should be DBZA or DBAZ? The paradigm established in \citet{sion1983} and \citet{wesemael1993} states that the spectral type is defined in order of the ``strongest'' optical spectral features, but no further definition is given as to what exactly that means.  It is ambiguous whether ``strongest'' refers to the  {\bf equivalent width} or the line {\bf depth}.  These comparisons can be substantially different depending on the instrument spectral resolution, especially for Ca {\small II} $\lambda$3933.663 {\AA} (CaK), which is typically the high-Z line with the largest EW in our temperature range (T$_{\rm eff}$ $<$ 18,000 K).  To illustrate this point, we list the CaK and H$\alpha$ line depths measured at both higher resolution (R $\sim$ 37,000) and lower resolution (R $\sim$ 1000), as well as their EWs in Table \ref{parameters}.  

If all we had were low resolution spectra, and if we chose to assign secondary spectral types by line depth, then four of the WDs would be DBZA and four DBAZ.  But then when those same WDs are observed at high resolution, according to line depth, the four previous DBAZs would all change to DBZAs.  
Instead, we decided to assign the spectral type according to EW:  DBAZ if EW(H$\alpha$) $>$ EW(CaK), and DBZA if EW(CaK) $>$ EW(H$\alpha$).  As long as spectra have sufficient signal-to-noise to detect a given line, EW measurements are essentially independent of the instrument resolution, and thus our choice of spectral type should be enduring.


\subsection{Stellar Parameters \label{abundmodel}}
The effects of additional opacity from the presence of hydrogen and heavier elements in the atmospheres of He-dominated WDs with effective temperatures ($T_{\rm eff}$) $<$ 20,000\,K have been well described \citep{dufour2007, dufour2010, coutu2019}. 

We follow an iterative procedure to obtain atmospheric parameters for each target. First, we get a rough estimate for $T_{\rm eff}$ and gravity (\,log $g$) by fitting photometry (typically Sloan Digital Sky Survey (SDSS), but PanSTARRS was used for EC22211$-$2525). We then fit the Ca {\small II} K (CaK) region and H$\alpha$ from low resolution spectra concurrently with SDSS {\it ugriz} photometry  \citep{alam2015} or PanSTARRS {\it grizy} photometry \citep{flewelling2020} and Gaia parallax \citep{gaiacollaboration2016,gaiacollaboration2021}. Where available (Table \ref{observations}) we use KAST spectra, otherwise we use SDSS spectra. Atmospheric structure calculations are then informed by the hydrogen abundance by number, n, (log $n$(H)/$n$(He)) and heavy element presence when scaling elements to the number abundance of Ca in a CI chondrite \citep{Lodders2019}.

We compared our fits to Gaia and GALEX photometry to confirm good agreement (see Figure \ref{SEDs}); standard de-reddening corrections were applied as described in \citet{coutu2019}. Our best-fit parameters are given in Table \ref{parameters}. We use these parameters to calculate the model atmospheres from which we produce synthetic spectra for each WD.


\begin{figure*}
\begin{center}
\includegraphics[width=6.4in]{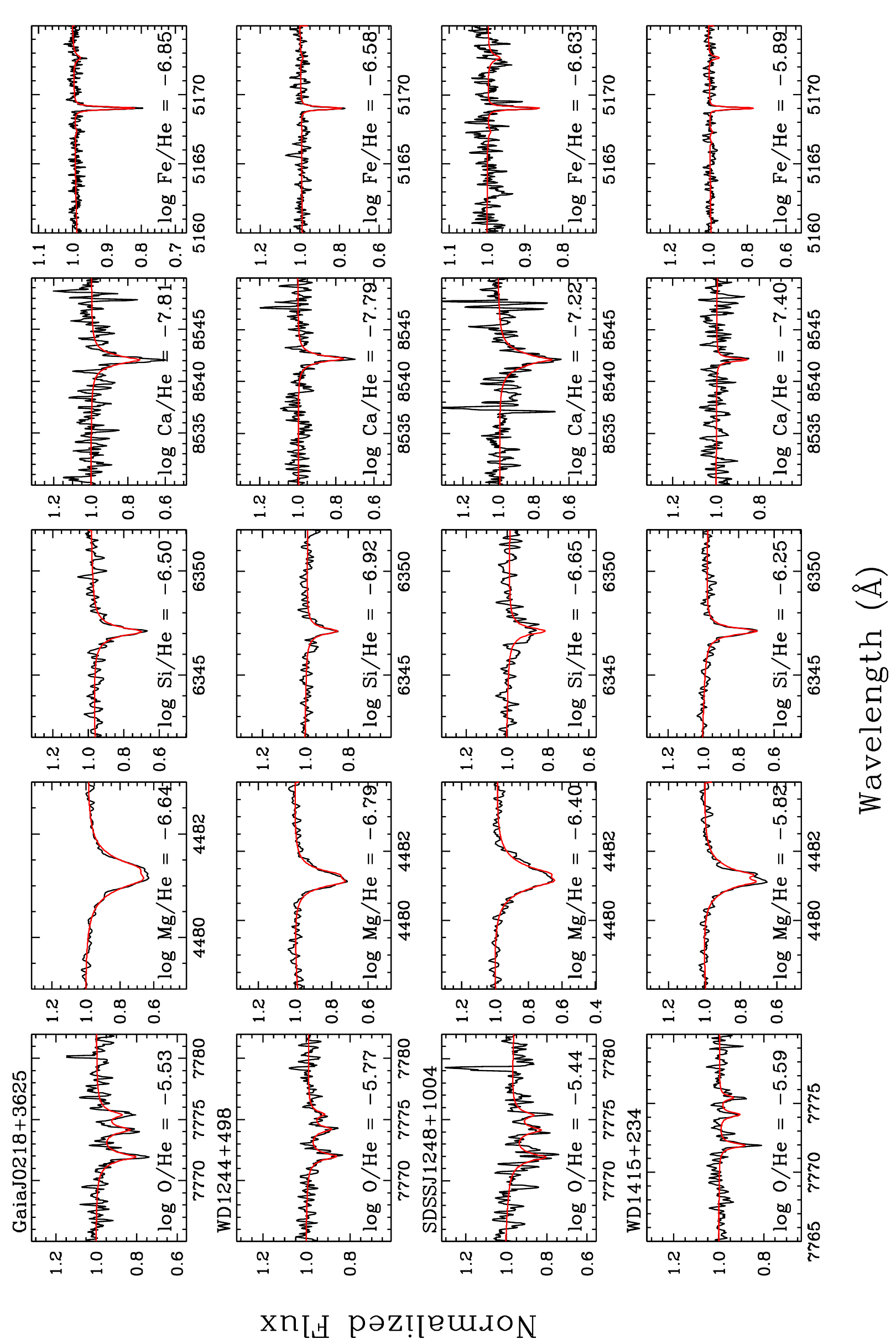}
\caption{Selected lines for each of the WDs in this study, displaying the detected O triplet and example lines for Mg, Si, Ca and Fe. Wavelengths are in air and shifted to the laboratory frame of rest. The red line is our best-fit model.}
\label{spectra1}
\end{center}
\end{figure*}

\renewcommand{\thefigure}{\arabic{figure} (cont.)}
\addtocounter{figure}{-1}
\begin{figure*}
\begin{center}
\includegraphics[width=6.4in]{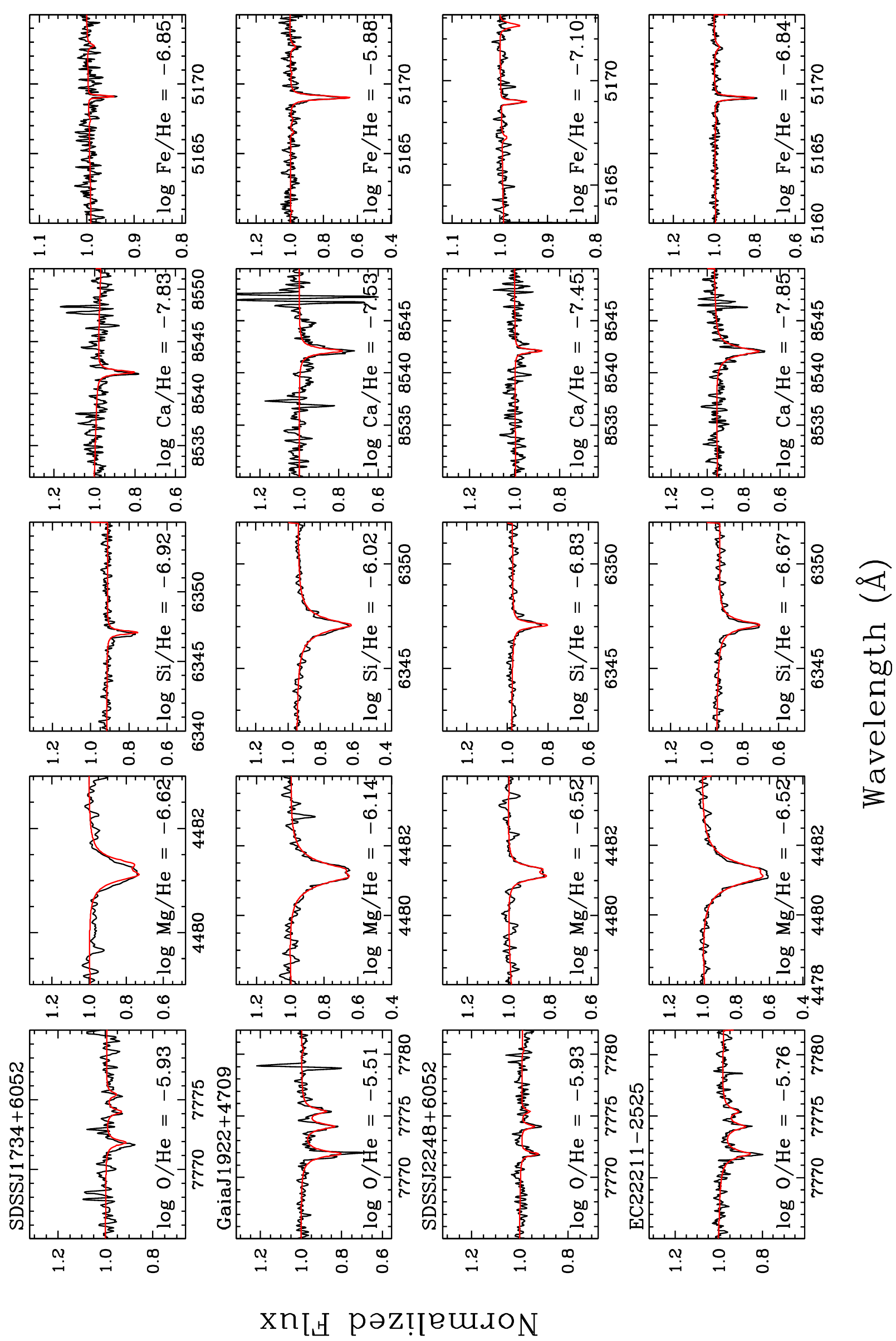}
\caption{Selected lines for each of the WDs in this study, displaying the detected O triplet and example lines for Mg, Si, Ca and Fe. Wavelengths are in air and shifted to the laboratory frame of rest. The red line is our best-fit model.}
\label{spectra2}
\end{center}
\end{figure*}
\renewcommand{\thefigure}{\arabic{figure}}

\begin{deluxetable*}{lccccc}
\tablecaption{Observed Atmospheric Elemental Abundances\label{abundances}}
\tablehead{	Name    &	log($n$(H)/$n$(He))	&	log($n$(Be)/$n$(He))	&   log($n$(O)/$n$(He))	&	log($n$(Na)/$n$(He)) & log($n$(Mg)/$n$(He))		}
\startdata																					
GaiaJ0218+3625 & $-6.03\pm	0.15$	&$<-11.0$	&	$	-5.53	\pm	0.15	$	&   $	-7.11	\pm	0.15	$   &$-6.64	\pm	0.15	$\\	
WD1244+498 & $-5.12\pm	0.15$	&$<-11.0$	&   $-5.77\pm	0.15	$	&   &	$-6.79\pm	0.15	$		\\
SDSSJ1248+1005 & $-5.18\pm	0.15$	&$<-11.0$	&	$-5.44\pm	0.15	$	&   &	$-6.40\pm	0.15	$		\\
WD1415+234  &   $-4.92\pm	0.15$	&$<-11.0$	&	$-5.59\pm	0.15	$	&   &	$-5.82\pm	0.17	$		\\
SDSSJ1734+6052 & $-4.76\pm	0.15$	&$<-10.3$	&$-5.93 \pm0.15$	&   &$-6.62 \pm0.15$	\\
GaiaJ1922+4709	& $-5.66\pm	0.15$	&$<-10.4$	&	$	-5.51	\pm	0.15	$	& &	$-6.14\pm	0.15$	\\
EC22211$-$2525	&$-5.56\pm	0.15$	&$<-11.0$	&   $	-5.76	\pm	0.15	$	&&	$	-6.52	\pm	0.15	$	\\	
SDSSJ2248+2632 & $-5.12\pm	0.15$	&$<-10.5$	&	$-5.94	\pm	0.15	$   &   &	$-6.52	\pm	0.15	$		\\
\hline						
Name	&	log($n$(Al)/$n$(He)) 	&	log($n$(Si)/$n$(He))&	log($n$(Ca)/$n$(He))	&	log($n$(Ti)/$n$(He))	&		 log($n$(Cr)/$n$(He))	\\
\hline
GaiaJ0218+3625 & $-7.3\pm 0.2$ &$	-6.50\pm0.15$&$-7.81\pm	0.21$&$	-9.43\pm0.15$&$-8.68\pm	0.15	$	\\	
WD1244+498 && $-6.92\pm	0.15	$&	$-7.79\pm	0.17	$	&	$-9.34\pm	0.15	$	&	$-8.78\pm	0.16$	\\
SDSSJ1248+1005 && $-6.65\pm	0.15	$&	$-7.22\pm	0.17	$	&	$-8.80\pm	0.15$	&	$-8.41\pm	0.15$	\\
WD1415+234  &&   $-6.25\pm	0.18	$&	$-7.40\pm	0.15	$	&	&	$-7.81\pm	0.15	$	\\
SDSSJ1734+6052 && $-6.93 \pm0.15$  &$-7.83 \pm0.17$	&	&	\\
GaiaJ1922+4709	&$-6.9\pm	0.2$	& $	-6.02	\pm	0.15$&	$	-7.53\pm0.15$	&$-9.05\pm	0.15$&$	-8.30\pm0.15$ \\
EC22211$-$2525	&$	-7.7	\pm	0.3	$	&   $	-6.67	\pm	0.15	$	&	$	-7.85	\pm	0.20	$	&$-9.60\pm0.15$&$-8.79\pm0.15$	\\
SDSSJ2248+2632 &	& $-6.83\pm	0.15	$&	$-7.45\pm	0.23	$	&	&	\\
\hline						
Name	&	 log($n$(Mn)/$n$(He))  &     log($n$(Fe)/$n$(He))   &     &&\\
\hline
GaiaJ0218+3625  &	$-8.84\pm	0.15$  & $	-6.85	\pm	0.15	$& &&\\	
WD1244+498 &	&   $-6.58\pm	0.15	$	&   &&\\
SDSSJ1248+1005 &&  $-6.63\pm	0.15	$	&&&\\
WD1415+234  &	&	$-5.89\pm	0.15	$   &&&\\
SDSSJ1734+6052 &	&$-6.85 \pm0.15$	&&&\\
GaiaJ1922+4709	&     &   $	-5.88	\pm	0.15	$ & 	&&\\
EC22211$-$2525	&	&	$	-6.84	\pm	0.15	$&&&\\
SDSSJ2248+2632 &	&   $-7.10\pm	0.27	$	&&&\\
\enddata
\tablecomments{Abundances by number, n, relative to He and uncertainties for each of the WDs in this work. Where statistical uncertainties are small ($<$0.15 dex), we conservatively set them to 0.15 dex. We have included upper limits on Be abundances, which demonstrate that Be is not detected at the greatly elevated levels seen in two WDs in \cite{klein2021}. We list observed lines used for these abundance determinations in Table \ref{EWs}.}			
\end{deluxetable*}

\subsection{Abundance Measurements \label{SpectralMeasurements}}
Over a series of multiple iterations, we fit these synthetic spectra to the HIRES data until we find a best-fit abundance solution for each element detected (Table \ref{abundances}). We show a sample of WD spectral lines for detections of O, Mg, Si, Fe, and Ca (Figure \ref{spectra1}).  In each panel our spectra are shown in black, and our best-fit model is overlain in red, and the numerical average abundance is given at the bottom of each panel. Our sample of eight WDs have clear detections of O (7772 \AA, multiplet), Mg (4481\AA, multiplet), Si (6347\AA), Fe (5169\AA), and Ca (3933\AA\ and 8542\AA), as well as other detected lines. 
Measured radial velocities (RVs) and a full listing of all detected lines with their EWs are given in the Appendix Tables \ref{rvs} and \ref{EWs}, respectively. We also discuss some detections of non-photospheric lines in the Appendix and Table \ref{rvs}.

Abundances are reported by number, n, relative to He along with uncertainties for each of the WDs in Table \ref{abundances}. Where elements are detected through multiple lines, we take the average abundance. Uncertainties are measured as the standard deviation where there are multiple lines of the same element. Systematic uncertainties, such as from uncertain atomic data \citep{vennes2011, gaensicke2012cos}, or other missing physics in atmosphere models \citep[e.g.][]{klein2020,cukanovaite2021} are difficult to quantify. Therefore, where only one line of an element is observed or where uncertainties are smaller than 0.15 dex, we conservatively set them to 0.15 dex.

\section{Discussion \label{discussion}}
\subsection{Accretion and Diffusion \label{steadystate}}
Three phases of accretion and diffusion of planetary debris onto a WD are commonly recognized in the literature: the buildup phase, sometimes referred to as an ``increasing" phase, the steady-state phase, and the settling, or ``decreasing" phase \citep[e.g.,][]{Dupuis_1992,Koester2009}. Though the specific nomenclature varies, the idea remains the same: as a single parent body accretes onto a WD, the observed pollution will first increase as material accumulates in the WD atmosphere. Then, as material begins to sink through the atmosphere, a steady state is eventually reached between accretion and diffusive settling. Steady state is achieved on a timescale comparable to a few e-folding times for settling. Once the parent body source is depleted, material ceases to accrete, and the observed pollution decreases commensurate with the settling times of the individual elements. 

The correction for this effect during steady-state accretion is straight forward $-$ element ratios are multiplied by the inverse ratio of settling timescales; see Equation 7 in \citet{Koester2009} and settling timescales in Table \ref{taus}.

While it is not clear which accretion state WDs exist in, ongoing accretion can be assumed for WDs with observed infrared excess, which emerge where circumstellar debris disks thermally reprocess the light from the star \citep{RN7}. EC22211-2525 is the only WD in the sample with detected infrared excess \citep{lai2021}, as can be seen in Figure \ref{SEDs}.

\begin{figure}
\begin{center}
 \includegraphics[width=3.4 in]{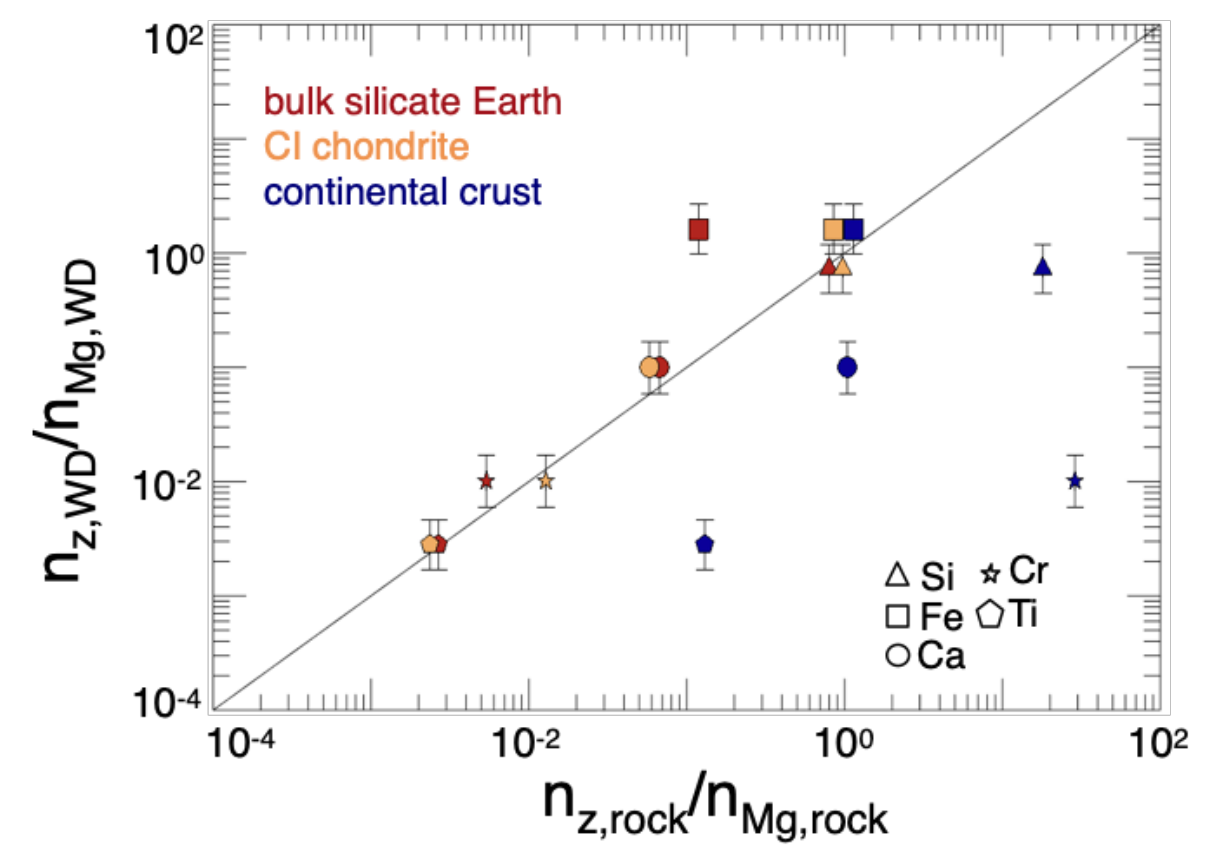}
 \caption{Element/magnesium atomic ratios, $z$/Mg, for the parent body accreted by WD1244$+$498, assuming an increasing phase, relative to $z$/Mg in various rocks found in our solar system. We compare the calculated parent body elemental abundances accreted by WD1244$+$498 to CI chondrite \citep[orange,][]{Lodders2019}, bulk silicate Earth (BSE) \citep[red,][]{RN28}, and the Earth's continental crust \citep[blue,][]{Rudnick2014}. The best match compositionally for the parent body accreting onto WD1244$+$498 is CI chondrite.}
   \label{EC_vs_rock}
\end{center}
\end{figure}

\begin{figure*}
\begin{center}
\includegraphics[width=185mm]{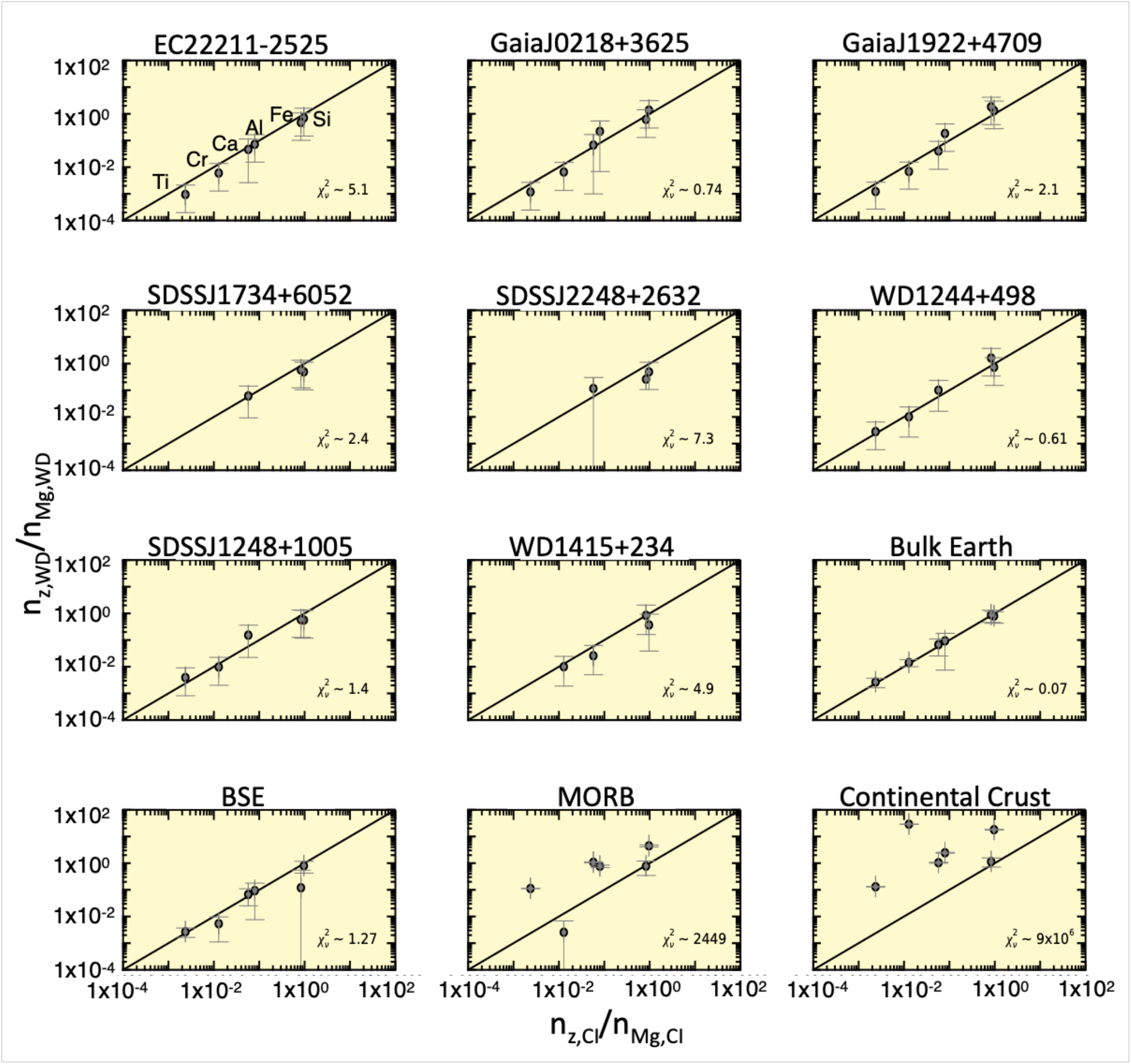}
\caption{One-to-one comparison of major and minor rock forming elements (n$_z$), ratioed to Mg (n$_{\rm Mg}$) and CI chondrite \citep{Lodders2019} for eight WDs. Abundances are from Table \ref{abundances}, representative of an increasing phase. Errors for WDs are propagated from model abundances and uncertainties using a Monte Carlo approach with a bootstrap of n=1. We report the goodness of fit using a reduced chi-square statistic, $\chi^2_\nu$, using the elements Si, Fe, Ca, Al, Cr, and Ti, where available for each WD (see text), displayed in the bottom right corner of each plot. Generally, the elemental abundances from WD data show good agreement with CI chondrites ($\chi^2_{\nu, {\rm crit.}} <$ 3.5-5.0, depending on which elements are used in the analysis, see text). For comparison, we calculate $\chi^2_\nu$ statistics for known compositions of Earth rocks (bulk Earth \citep{RN28}, bulk silicate Earth \citep[BSE,][]{RN28}, Mid-Ocean Ridge Basalt \citep[MORB,][]{gale2013}, and the Earth's continental crust \citep{Rudnick2014}) compared to CI chondrite. Bulk Earth and bulk silicate Earth are in good agreement with CI chondrite, revealing that WD-sized errors in the elements used (Ti, Cr, Ca, Al, Fe, and Si) are unable to distinguish between the two compositions in the data.}
\label{1to1}
\end{center}
\end{figure*}

\begin{figure*}
\begin{center}
\includegraphics[width=185mm]{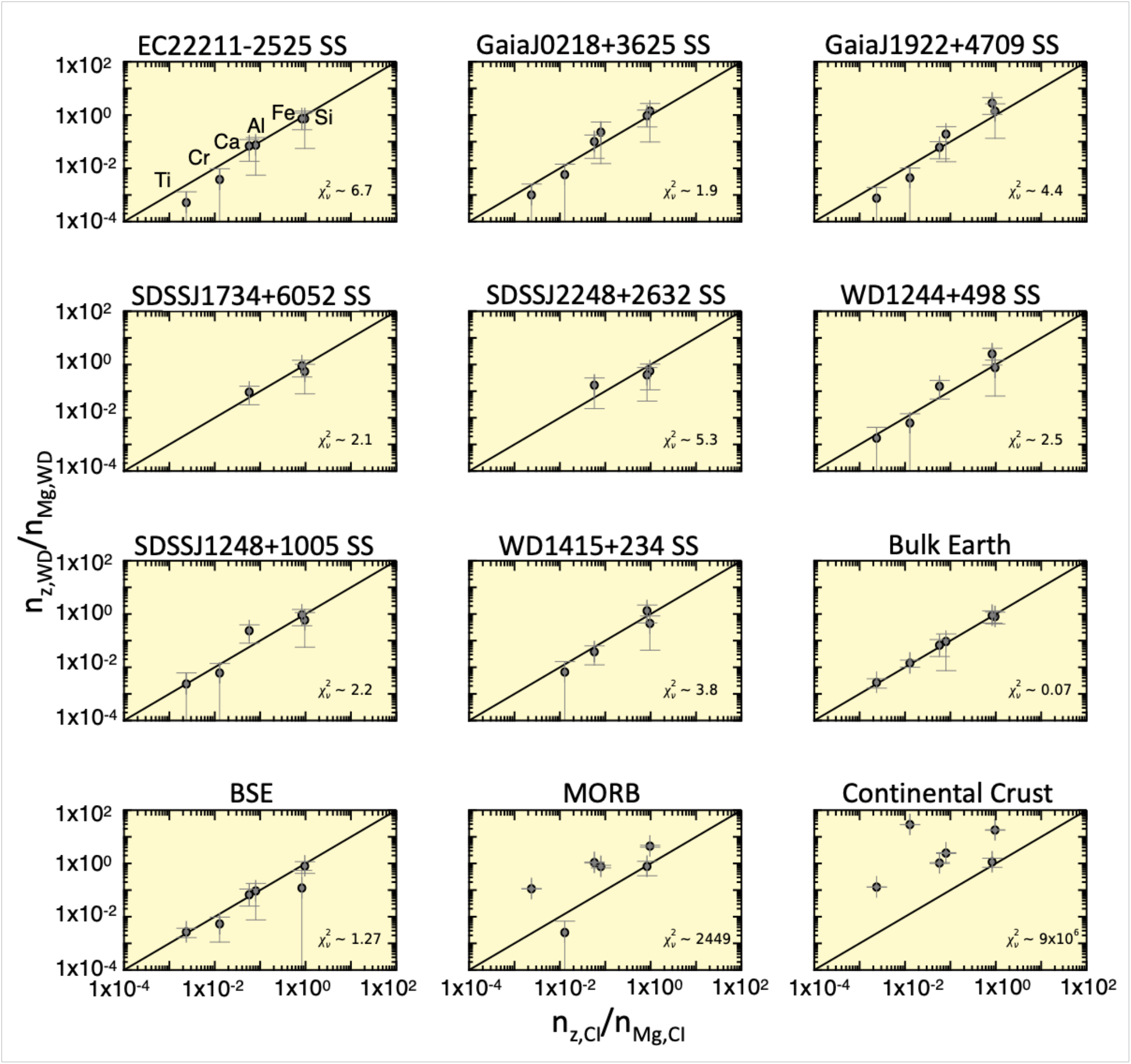}
\caption{As in Figure \ref{1to1}, but assuming steady-state phase (SS) compositions for the eight WDs presented.}
\label{SS1to1}
\end{center}
\end{figure*}

\subsection{Abundance Pattern \label{abundancepattern}}
For each of the WDs in this study we compared the observed abundances of rock-forming elements (Mg, Al, Si, Ca, Ti, Cr, and Fe) to those of typical rocky compositions in the solar system (CI chondrite, bulk silicate Earth, and continental crust). In general, the best fit is to CI chondrite. In Figure \ref{EC_vs_rock} we illustrate this result using the composition of the parent body polluting WD1244$+$498 as an example. The parent body is comparable to CI chondrite, as indicated by the close agreement of chondritic abundances (orange symbols) to the 1:1 line in Figure \ref{EC_vs_rock}. Indeed, each element agrees with chondritic compositions within a factor of 2. 

Motivated by Figure \ref{EC_vs_rock}, we statistically evaluate the hypothesis that the parent bodies being accreted by these eight WDs were approximately chondritic in composition. Similar to \citet{xu2013gd362pg1225}, \citet{Swan_2019}, and \citet{Doyle2021}, we compare the goodness-of-fit for rock-forming elemental abundances observed in each WD to the known composition of CI chondrites using the reduced chi-square statistic, $\chi^2_\nu$ (Figure \ref{1to1}). We calculate $\chi^2_\nu$ using the elements Al, Si, Ca, Ti, Cr, and Fe, where available for each WD. Oxygen is excluded due to its correlation with the other rock-forming elements (see Section \ref{oxygen}). Additionally, because  we ratio elements to Mg, Mg is not an independent observation for this calculation and is therefore excluded. The data points and their uncertainties shown in Figure \ref{1to1} represent propagated uncertainties using a Monte Carlo approach with a bootstrap of n=1. 

The parameter $\alpha$ represents a probability of obtaining $\chi^2_\nu$ values greater than the observed value by chance. Convention suggests that the threshold to reject the hypothesis that the data are consistent with a CI composition is 5\% or better, or $\alpha < 0.05$ ($\alpha \sim 0.4$ for $\chi^2_\nu =1$). Due to the relatively small number of data points per star, and their uncertainties, the value of $\chi^2_\nu$ is also uncertain, which can be accounted for using the approach of  \cite{andrae2010} in which the uncertainty in $\chi^2_\nu$ is $\sigma \sim \sqrt{2/N}$ for $N$ data points. Based on a threshold for $\alpha = 0.05$ and a 2$\sigma$ error for $\chi^2_\nu$, we define a critical value, $\chi^2_{\nu, {\rm crit}}$, as the reduced chi-square value corresponding to $\alpha = 0.05 + 2\sigma$. Based on these critical values, ranging from $3.5$ to $5.0$, depending on the number of elements involved, the relative elemental abundances for the polluted WDs examined here are in good agreement with CI chondrites, with 5 of the 8 WDs having values for $\chi^2_\nu$ less than the associated critical values. The remaining WDs have values for $\chi^2_{\nu}$ of $5.08$, $7.3$, and $4.9$, making their fits to CI tentative. For context, we also calculate $\chi^2_\nu$ for the bulk Earth, bulk silicate Earth, and terrestrial crustal rocks compared to CI chondrite, where we assume errors equal to the average WD error for each element ratioed to Mg, $n_z/n_{\rm M_g}$. Note that bulk Earth and bulk silicate Earth (BSE) are indistinguishable from CI chondrite in this analysis using uncertainties associated with the WD observations of Mg, Al, Si, Ca, Ti, Cr, and Fe. The compositions of continental and oceanic crust, the latter represented by Mid-Ocean Ridge Basalt (MORB), are readily distinguished from CI chondrite in major elements using WD uncertainties (Figure \ref{1to1}). We see no evidence for crust-like compositions among the eight polluted WDs considered here.

In the examples presented above, we used the observed elemental ratios with no corrections for settling times. This tacitly assumes that the parent body accretion is in the buildup phase. We calculate the same  $\chi^2_\nu$ statistic to assess the goodness-of-fit for these WDs relative to the CI elemental ratios assuming the WDs are accreting material in a steady-state phase (Figure \ref{SS1to1}). Steady state is often assumed for WDs in which heavy element settling times are relatively short. Under this assumption, we find that for 3 of the 8 WDs, the $\chi^2_\nu$ values relative to CI chondrite indicate better agreement with CI chondrite than for the buildup-phase assumption. However, with the steady-state assumption, still 5 of the 8 WDs are indistinguishable from CI chondrites ($\chi^2_\nu < \chi^2_{\nu, crit.}$). Therefore, regardless of whether these polluted WDs are assumed to be in the buildup phase or in steady state, they appear to be accreting bodies that are chondritic, or approximately chondritic, in composition. We note that for GaiaJ0218$+$3625 (irrespective of accretion phase) the abundance of Na/Mg is $\simeq$6$\times$ the chondritic ratio. There is likely more work to be done in future analysis of GaiaJ0218$+$3625, but this particular enhanced relative abundance is not sufficient alone to reject the assessment that overall, the accreted bodies of this sample are broadly chondritic.

\subsection{Parent Body Size \label{parentbody}}
In order to estimate  parent body sizes, we calculate the minimum masses of the parent bodies accreting onto these eight WDs as the sum of the masses of all heavy elements in the convection zone (CVZ). We convert number abundance ratios from Table \ref{abundances} to mass ratios and multiply by the mass of the convection zone, computed from evolution models from the Montreal White Dwarf Database \citep[MWDD;][]{dufour2007}\footnote{http://dev.montrealwhitedwarfdatabase.org/evolution.html}. We find minimum masses that range from 2.8 $\times$ 10$^{21} -$ 9.0 $\times$ 10$^{22}$ g. These masses are consistent with some of the most massive asteroids in the solar system ($\thicksim$8 Flora $-$ 10 Hygiea) and some of the mid-sized moons in the solar system ($\thicksim$ Neptune's Larissa $-$ Saturn's Enceladus). The immensity of these minima for parent body masses supports the conclusion that only the most massive of polluting objects will be observable in WDs \citep{Trierweiler2022}. Mass fluxes onto the WD atmosphere can be obtained by assuming steady state between accretion and settling. For this we use the CVZ pollution masses and settling times from Table \ref{taus}. The derived fluxes range from 1.4 $\times$ 10$^8 -$ 8.5 $\times$ 10$^9$ g s$^{-1}$, typical for polluted WDs under similar assumptions \citep[e.g.][]{rafikov2011mnras,farihi2012gas,Wyatt_2014,Xu_2019} and would result in parent body masses that range from 2.1$\times$ 10$^{21} -$ 1.4 $\times$ 10$^{23}$ g, assuming accretion from a disk is sustained for roughly 5$\times$10$^5$ yr \citep{girven2012}.

\subsection{Oxygen and Oxidation State \label{oxygen}}
We evaluate the oxidation state of the parent bodies accreting onto each WD by following the prescription introduced by \citet{Doyle_2019} and improved in \citet{doyle2020}. We use the ratio of $\rm O_{rem}$/Fe, where $\rm O_{rem}$ is the O remaining after assigning O to Mg, Si, Ca, and Al to form the oxides MgO, SiO$_2$, CaO, and Al$_2$O$_3$, as an indicator for whether a WD will yield a recoverable oxygen fugacity (\DIW, see discussion below for complete definition for this parameter) value and error bounds. We calculate $\rm O_{rem}$ relative to Fe as:

\begin{equation}
\frac{\rm{O}_{rem}}{\rm Fe} = \frac{\rm O}{\rm Fe} -\frac{\rm Mg}{\rm Fe} -  2\frac{\rm Si}{\rm Fe} - \frac{3}{2}\frac{\rm Al}{\rm Fe} - \frac{\rm Ca}{\rm Fe}.
   \label{Oxs_equation}
\end{equation}

\noindent For an ideal rock, in which Fe exists as ferrous iron (effective charge of 2+), the value of $\rm O_{rem}$/Fe should be unity. Where $\rm O_{rem}/Fe >1$, an oxygen excess exists, suggesting an additional source for oxygen, often due to accretion of oxygen-bearing volatiles such as H$_2$O from the parent body (we exclude the effect of Fe$^{3+}$ here, present as the oxide Fe$_2$O$_3$, under the assumption that the ferric iron will be relatively minor, $< 10\%$ of all Fe, as it is in most solar-system rocks). Six of the eight WDs in this study have observed oxygen excesses implying water-rich bodies ($\rm O_{rem}/Fe >1$; Table \ref{fo2}). Of the six WDs with oxygen excesses, five have an observed amount of H that can account for the excess oxygen assuming a buildup phase. Large abundances of H in helium-dominated WDs are either from primordial H (prior to the DA-to-DB evolution, \citealt{rolland2020}) or due to the accumulation of H throughout accretion events, as H floats on the atmospheric surface \citep{gentilefusillo2017,izquierdo2021}. Notably, a steady-state approximation decreases, but does not entirely remove, the oxygen excesses (Table \ref{fo2}). 

The level of oxidation in a geochemical system is described as the non-ideal partial pressure of O$_2$, or oxygen fugacity ($f_{\rm O_2}$), and has implications for the geochemistry and geophysics of rocky bodies. In the planet formation regime, oxygen fugacities are often compared with that defined by the equilibrium reaction between metallic iron (Fe) and FeO, which in mineral form is w$\ddot{\text{u}}$stite (FeO):

\begin{equation}
\text{Fe} + \frac{1}{2} \text{O}_2 \rightleftharpoons \text{FeO} .
\label{feo}
\end{equation}

\begin{table*}
\caption{Oxidation States Determined from WD Data in this Study \label{fo2}}
\begin{center}
\begin{tabular}{lccc}
\hline
\hline
Name    &	\DIW	&	$\rm{O}_{\rm{rem}}/\rm{Fe}$	&	$\rm{O}_{\rm{rem}}/\rm{Fe}$ (steady)\\
\hline
GaiaJ0218+3625  & $	-1.29	^{+	0.27}_{-0.37}$	&   $	14.16	^{+11.02}_{- 7.52}$ &	$	9.28	^{+	8.68}_{- 6.68}	$\\	
WD1244+498      & $-0.54^{+	0.17}_{-0.26}	$	&	$4.69^{+ 3.53}_{- 2.34}	$		&	$	3.01	^{+	2.65}_{- 2.03}	$\\
SDSSJ1248+1005  & $-1.09^{+ 0.24}_{-0.34}	$	&	$11.23^{+ 8.58}_{-5.68}	$		&	$	7.11	^{+	6.40}_{- 4.90}	$\\
WD1415+234      & $<-0.87	$	                &	$-0.18^{+	0.91}_{-0.94}	$	&	$	-0.30	^{+	0.71}_{- 0.79}	$\\
SDSSJ1734+6052  & $-1.04 ^{+0.26}_{-0.43}$	    &   $4.62 ^{+4.13}_{-2.97}$	        &	$	2.72	^{+	3.24}_{- 2.64}	$\\
GaiaJ1922+4709	& $	-1.78	^{+	1.17}_{}	$	&	$	0.17	^{+	1.01}_{- 0.97} $&	$	0.03	^{+	0.89}_{- 0.88}	$\\
EC22211$-$2525	& $	-1.25	^{+	0.28}_{- 0.44}$	&	$	6.73	^{+	6.07}_{- 4.36}$	&	$	4.27	^{+	4.82}_{- 3.89}	$\\	
SDSSJ2248+2632  & $-1.74	^{+	0.49}_{}$       &	$5.01	^{+	8.83}_{-5.48}	$	&	$	2.44	^{+	5.77}_{- 4.73}	$\\
\hline
\end{tabular}
\end{center}
\tablecomments{Calculated \DIW and remaining O relative to Fe, along with error bounds for the WDs in this study. $\rm{O}_{\rm{rem}}/\rm{Fe}$ for an ideal rock should be unity, and variations from this value are due to oxygen either in excess or shortage of that required to form MgO, SiO$_2$, CaO, and FeO. Measurement uncertainties are propagated using a Monte Carlo approach with a bootstrap of n=1; see Section \ref{oxygen} for discussion about absent lower error bounds for \DIW. Generally, a steady-state assumption reduces the remaining oxygen, but does not entirely remove the excess, implying that the 6 WDs with oxygen excesses in the steady-state calculation have some amount of oxygen-bearing volatiles, such as H$_2$O ice, in the parent body.}
\end{table*}

\noindent This iron-w$\ddot{\text{u}}$stite (IW) reference reaction assumes pure Fe metal and FeO oxide. By reporting $f_{\rm O_2}$ of a rock to a reference reaction such as Equation \ref{feo}, the thermodynamics simplifies to a ratio of activities, or mole fractions (see Appendix in \citet{Doyle_2019} for a full derivation). The intrinsic oxygen fugacity of a rock or rocky body can thus be described relative to that for the IW reference, such that

\begin{equation}
\Delta \text{IW} \equiv \text{log} \left( f_{\rm O_2} \right)_{\rm rock} - \text{log} \left( f_{\rm O_2} \right)_{\rm IW} = 2 \text{log} \left( \frac{{x_{\rm FeO}^{\rm rock}}}{{x_{\rm Fe}^{\rm metal}}}  \right).
  \label{DIW}
\end{equation}

\noindent This simplification results in an equation for \DIW that depends solely on the mole fraction of FeO in the rock ($x_{\rm FeO}^{\rm rock}$) and the mole fraction of Fe in the metal ($x_{\rm Fe}^{\rm metal}$).

Where $\rm O_{rem}/Fe <1$, a dearth of oxygen exists, suggesting iron is present in the form of Fe metal. Of the eight WDs reported in Table \ref{fo2}, three have lower bounds with values for $\rm O_{rem}/Fe <1$ (WD1415$+$234, GaiaJ1922$+$4709, and SDSSJ2248$+$2632). In such cases, lower bounds on the level of oxidation, measured as oxygen fugacity, cannot be obtained.

\begin{figure}
\begin{center}
\includegraphics[width=3.25 in]{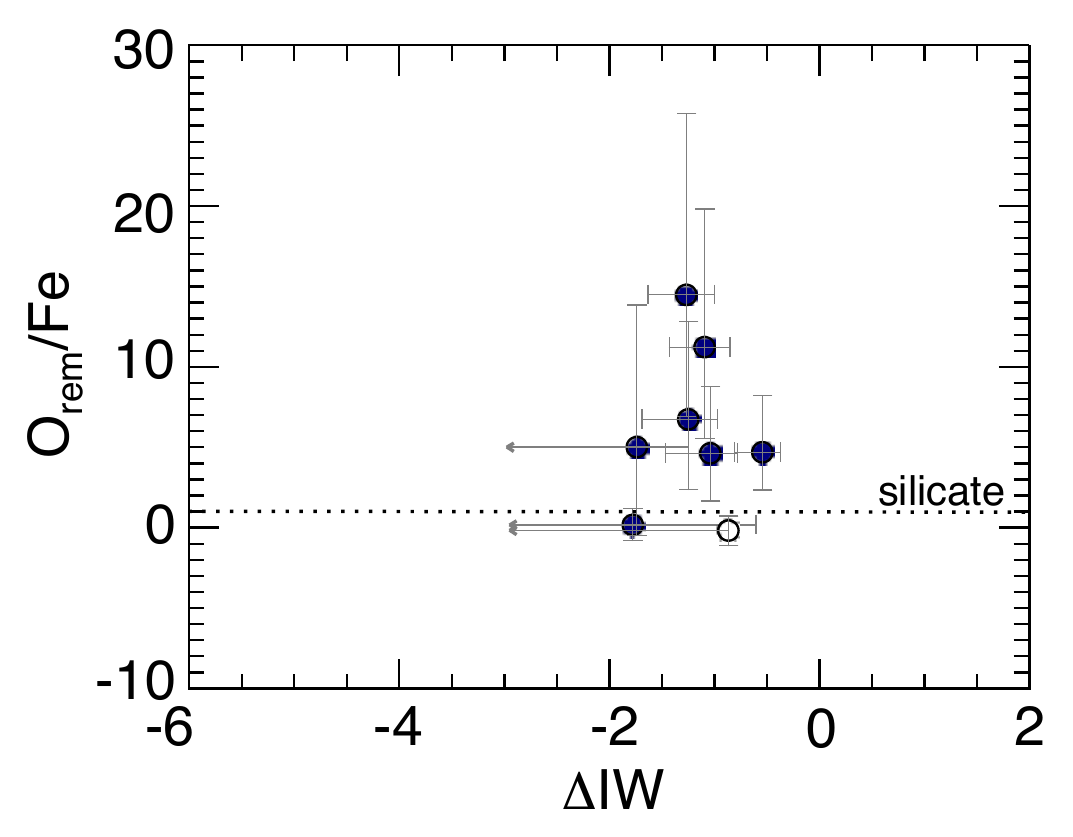}
\caption{\DIW vs $\rm O_{rem}$/Fe. The value of $\rm O_{rem}$/Fe should be unity for ideal rocks, represented as the dotted silicate line. Where $\rm O_{rem}/Fe >1$, an oxygen excess exists, and where $\rm O_{rem}/Fe <1$, a dearth of oxygen exists. Where errors allow $\rm O_{rem}/Fe <1$, lower bounds in \DIW cannot be obtained. Such is the case for three WDs in this study (GaiaJ1922$+$4709, SDSSJ2248$+$2632, and WD1415$+$234). These three WDs are those in the Figure where lower error bounds in $\rm O_{rem}/Fe$ plot below the ideal value for silicates.} Of these three, WD1415+234 is represented with an open circle, because the median is also unrecoverable.
\label{DIWvsOxs}
\end{center}
\end{figure}

As in \citet{Doyle_2019} and \citet{doyle2020}, we use the oxides SiO$_2$, MgO, FeO, CaO and Al$_2$O$_3$ to characterize the chemical composition of the accreting rocks. Where Al is not observed, we assume a chondritic Al/Ca ratio and set uncertainties equal to 0.3 dex. Using oxides ensures charge balance and provides a means of tracking oxygen that was in the form of rock. We first assign oxygen to Mg, Si, and Ca to form these oxides, and then we assign the remaining oxygen, O$_{\rm rem}$, to Fe to form FeO. In this way, we can assess what portion of Fe can be paired with O and is presumed to have existed as FeO in the rock ($x_{\rm FeO}^{\rm rock}$) versus what portion of Fe existed as Fe metal (i.e. where there is a deficit of O). For application of Equation \ref{DIW} we set $x_{\rm Fe}^{\rm metal} =$ 0.85, consistent with estimates for Fe metal in the core of differentiated bodies from our solar system. We propagate measurement uncertainties for the polluted WDs using a Monte Carlo approach with a bootstrap of n=1. We report our calculated \DIW values in Table \ref{fo2}. 

In our solar system, most rocky bodies are oxidized relative to a hydrogen-rich solar gas (\DIW $= -6$), with \DIW values greater than $-3$, corresponding to $x_{\rm FeO}^{\rm rock} > 0.025$. Only Mercury and enstatite chondrites are ``reduced" (\DIW $< -3$; $x_{\rm FeO}^{\rm rock} < 0.025$). In general, the WDs in this study have \DIW values similar to chondrites, consistent with their chondritic bulk chemistry (Figures \ref{1to1} and \ref{SS1to1}). However, there are two WDs in this study for which lower bounds on \DIW cannot be obtained (GaiaJ1922$+$4709 and SDSSJ2248$+$2632), and one for which neither a median nor a lower bound can be obtained (WD1415$+$234). Situations like these arise where negative $x_{\rm FeO}^{\rm rock}$ values are a significant fraction of the Monte Carlo draws for error propagation. This in turn comes about where there is either a relative scarcity of oxygen relative to the propagated errors or abundance uncertainties are large (refer to Section 2.3 and Figure 3 in \citet{doyle2020} for a more detailed discussion).

Therefore, the calculation of $\rm O_{rem}$/Fe is a good indicator for whether a WD will yield a recoverable \DIW value and error bounds. Indeed, the same three WDs that have lower bounds with negative values for $\rm O_{rem}$/Fe have unrecoverable lower error bounds for \DIW (Figure \ref{DIWvsOxs}). It is worth noting that one of these WDs, GaiaJ1922$+$4709, is that with the least good fit to CI chondrite, based on $\chi^2_\nu$ statistics presented in Figure \ref{1to1}. It is also worth noting that one of these WDs, SDSSJ2248$+$2632, has a median value for $\rm O_{rem}$/Fe that indicates excess oxygen, but large uncertainties for Fe (Table \ref{abundances}). Indeed, it is possible that the parent bodies accreting onto these WDs had less FeO in the rocky portion of the body and were more reduced than CI chondrite. While these WDs have oxidation states that are less constrained, the median values for \DIW calculated for this subset of polluted WDs generally adds to the increasing quantity of chondrite-like parent bodies accreting onto WDs in both bulk composition and degree of oxidation.

\section{Conclusions  \label{conclusions}}
In this work we present observations for eight heavily polluted DB white dwarfs and relative elemental abundances for the rocky parent bodies that accreted onto them. All of the WDs in this data set required new designations or updates of spectral types. In a step towards some needed clarification to the spectral classification system,  we measured and ordered the ``strongest'' spectral features according to equivalent widths (not line depths).  That determined our assignment of spectral types as DBAZ or DBZA.


We assembled our dataset by comparing GALEX colors to $T_{\rm eff}$ for white dwarf candidates presented in \citet{gentilefusillo2019}, as well as from known polluted DB white dwarfs. This comparison reveals a distinct dichotomy between DA and DB white dwarfs, which we used to target DB white dwarfs to search for those that are heavily polluted. The white dwarfs presented here were chosen due to their detections of all four major rock-forming elements (O, Mg, Si, Fe). Through this work, we have increased the sample of known oxygen-bearing white dwarfs polluted by rocky parent bodies by $\thicksim$ 33\% \footnote{see also note added in proof}.

We assessed the  bulk compositions and oxidation states of the accreting bodies, and find that they are indistinguishable from chondritic in composition. This adds to the growing body of evidence suggesting that extrasolar rocky bodies closely resemble those in our solar system, and do not, as a whole, yield unusual or unique compositions. This result is not dependent on assumptions of an increasing phase versus a steady-state phase of accretion. 

Six of the eight white dwarfs in this study have observed oxygen excesses implying volatiles, in various abundances, in the parent bodies (a trait shared by CI chondrites). Generally, the oxidation states of these parent bodies also corroborate the conclusion that the accreting bodies are chondritic. Three exceptions exist in which oxidation states are less constrained and could be more reduced than chondritic (lower oxygen fugacity values), and one of these white dwarfs (GaiaJ1922$+$4709) is the same WD that obtains the least good fit to CI chondrite. This result is in accordance with the assessment that perhaps 1/4 of polluted white dwarfs may be consistent with more reduced parent bodies that cannot be identified by use of this method \citep{doyle2020}. Overall, our results are consistent with the emerging view that extrasolar rocks across the solar neighborhood are broadly similar to rocky bodies in our solar system. 

\acknowledgements

This work was supported by NASA 2XRP grant No. 80NSSC20K0270 to EDY. C.M.\ and B.Z.\ acknowledge support from NSF grants SPG-1826583 and SPG-1826550. S. Xu is supported by the international Gemini Observatory, a program of NSF’s NOIRLab, which is managed by the Association of Universities for Research in Astronomy (AURA) under a cooperative agreement with the National Science Foundation, on behalf of the Gemini partnership of Argentina, Brazil, Canada, Chile, the Republic of Korea, and the United States of America. 

The authors thank Simon Blouin (University of Victoria) for helpful discussions about abundance modeling, and Jay Farihi (University College London) for helpful discussions regarding WD spectral type classifications. We also thank the anonymous reviewer for their comments, which improved the manuscript.

Much of the data presented herein were obtained at the W. M. Keck Observatory, which is operated as a scientific partnership among the California Institute of Technology, the University of California and the National Aeronautics and Space Administration. The Observatory was made possible by the generous financial support of the W. M. Keck Foundation. The authors wish to recognize and acknowledge the very significant cultural role and reverence that the summit of Maunakea has always had within the indigenous Hawaiian community.  We are most fortunate to have the opportunity to conduct observations from this mountain.

Similarly, we acknowledge that Lick Observatory resides on land traditionally inhabited by the Muwekma Ohlone Tribe of Native Americans. Research at Lick Observatory is partially supported by a generous gift from Google.

This paper includes data gathered with the 6.5 meter Magellan Telescopes located at Las Campanas Observatory, Chile.

This work has made use of data from the European Space Agency (ESA) mission Gaia (https://www.cosmos. esa.int/gaia), processed by the Gaia Data Processing and Analysis Consortium (DPAC). Funding for the DPAC has been provided by national institutions, in particular the institutions participating in the Gaia Multilateral Agreement. This research has made use of NASA’s Astrophysics Data System, the SIMBAD database, and the VizieR service. This research has made use of IRAF. IRAF is distributed by the National Optical Astronomy Observatory, which is operated by the Association of Universities for Research in Astronomy (AURA) under a cooperative agreement with the National Science Foundation.

The following atomic spectral line databases were consulted: Vienna Atomic Line Database (VALD), Kurucz (1995, R. L. Kurucz and B. Bell, CD-ROM No. 23, Cambridge, MA: Smithsonian Astrophysical Observatory), NIST Standard Reference Database 78, and van Hoof (2018). 

Facilities: Shane (Kast), Keck I (HIRES), Keck II (ESI), Magellan (MagE)

{\it Note added in proof.}  Contemporaneous with this paper, \cite{izquierdo2023} reported oxygen detections in ten polluted WDs, though parent body composition analyses have not yet been carried out.

\clearpage
\appendix{\label{appendix}}
\setcounter{table}{0}
\renewcommand{\thetable}{A\arabic{table}}
\setcounter{figure}{0}
\renewcommand{\thefigure}{A\arabic{figure}}
This appendix presents details of spectral line measurements, broadband spectral energy distributions (SEDs), diffusion timescales and accretion rates.  Radial velocites (RVs) are given in Table \ref{rvs}, equivalent widths (EWs) are given in Table \ref{EWs}, SEDs are displayed in Figure \ref{SEDs}, and accretion-diffusion data are reported in Table \ref{taus}. Equivalent widths were measured by profile fitting using IRAF's {\it splot} task, and RVs were calculated as Doppler shifts of the measured line centers relative to laboratory wavelengths \citep[see][]{klein2021}.  


Half of the stars in this sample (WD1415$+$234, SDSSJ1734$+$6052, GaiaJ1922$+$4709, SDSSJ2248$+$2632) display absorption lines of the Na {\small I} resonance doublet $\lambda$5889.951/5895.924\,{\AA} (NaD) with RVs which are significantly blue-shifted from the photospheric averages based on many photospheric lines (see Table \ref{rvs}).  In some stars non-photospheric Ca {\small II} $\lambda$3933.663 {\AA} (CaK) features are also observed. Based on results from \citet[][\url http://lism.wesleyan.edu/LISMdynamics.html]{redfieldlinsky2008} and \citet{welsh2010}, it is probably the case that WD1415+234, SDSSJ1734+6052, and GaiaJ1922+4709 host interstellar medium (ISM) features.

On the other hand, if the non-photospheric RV is blue-shifted from the photospheric RV by an amount equal to or somewhat less than the gravitational redshift of the WD, then it could be that the non-photospheric absorption is occurring in CS gas (co-moving with the WD, but not fully in its photospheric gravitational well).
Referring to Table \ref{rvs}, and considering an uncertainty range of 3 km s$^{-1}$ in gravitational redshift plus 2 km s$^{-1}$ in photospheric RV, a CS origin is reasonable for only two WDs: SDSSJ1734+6052 and SDSSJ2248+2632. However,  we can not rule out the possibility that absorption may be due to ISM material (especially at distances $>$ 80 pc; e.g., see Figure 7 of \citealt{welsh2010}) or could even possibly have some association with accretion-related outflows. 

Unlike the four aforementioned WDs, the NaD RV in GaiaJ0218+3625 agrees exactly with the average photospheric RV. There may be a  slight chance that an ISM cloud has the unusually high RV of 39 km s$^{-1}$ (e.g., \citealt{redfieldlinsky2008} and \citealt{welsh2010}) and is coincidentally the same as the WD RV. We think this unlikely, and deem the Na line in GaiaJ0218+3625 to originate in the WD photosphere and be associated with the polluting parent body. 

\begin{table*}
\caption{Radial Velocities \label{rvs}}
\begin{center}
\hspace*{-3.7cm}
\begin{tabular}{lccccrccccccc}
\hline
\hline
	&	& \multicolumn{3}{c}{photospheric lines} & sys- & \multicolumn{4}{c}{additional lines}	 & vel diff & grav & origin \\
\cmidrule(lr){3-5}  \cmidrule(lr){7-10}
WD Name	&	D	&	\# of	&	avg	&	std	&	temic	&	NaD$_2$	&	NaD$_2$	&	CaK	&	CaK	&	from	&	red-	&	of	\\
	&		&	lines	&	RV	&	dev	&	RV	&	EW	&	RV	&	EW	&	RV	&	phot	&	shift	&	extra	\\
	&	(pc)	&		&	(km s$^{-1}$)	&	(km s$^{-1}$)	&	(km s$^{-1}$)	&	(m\AA)	&	(km s$^{-1}$)	&	(m\AA)	&	(km s$^{-1}$)	&	(km s$^{-1}$)	&	(km s$^{-1}$)	&	lines	\\
\hline																									
GaiaJ0218+3625	&	116	&	83	&	39.5	&	1.6	&	16.1	&	45	&	39.2	&		&		&	0.2	&	23.4	&	phot	\\
WD1244+498	&	120	&	46	&	39.1	&	2.5	&	11.1	&		&		&		&		&		&	28.0	&		\\
SDSSJ1248+1005	&	164	&	62	&	41.0	&	1.9	&	5.6	&		&		&		&		&		&	35.4	&		\\
WD1415+234	&	127	&	57	&	34.1	&	1.5	&	-4.7	&	36	&	-15.8	&		&		&	49.9	&	38.8	&	ISM	\\
SDSSJ1734+6052	&	150	&	25	&	17.7	&	1.9	&	-13.9	&	69	&	-31.3	&	9	&	-28.1	&	47.8	&	31.6	&	ISM	\\
	&		&		&		&		&		&	60	&	-21.1	&	7	&	-18.7	&	38.2	&	31.6	&	ISM or CS?	\\
GaiaJ1922+4709	&	127	&	104	&	44.8	&	1.4	&	17.7	&	33	&	-22.6	&		&		&	67.8	&	27.1	&	ISM	\\
	&		&		&		&		&		&		&		&	37	&	-16.1	&	60.9	&	27.1	&	ISM	\\
EC22211-2525	&	109	&	88	&	47.1	&	1.6	&	22.6	&		&		&		&		&		&	24.5	&		\\
SDSSJ2248+2632	&	123	&	16	&	21.2	&	1.6	&	-9.4	&	54	&	-9.1	&		&		&	29.9	&	30.6	&	ISM or CS?	\\
\hline
\end{tabular}
\end{center}
\tablecomments{Radial velocities are in the heliocentric frame of rest.  ``avg'' RV is the average of the set of observed high-Z line RVs (includes gravitational redshift); ``std dev'' is the standard deviation of that set, which can be taken to be a measure of the uncertainty on the average; ``systemic'' RV is the kinematic portion of the RV (gravitational redshift subtracted); ``vel diff from phot'' = avg RV $-$ NaD and/or CaK RV; ``phot'' = photosphere. CaK measurements in this table are of secondary features associated with the CaK wavelength (3933.663\AA\ in air) that are velocity shifted from the photospheric CaK lines (reported in Table \ref{EWs}).  NaD$_2$ =  $\lambda$5889.951{\AA}, the stronger line of the Na {\small I} resonance doublet. SDSSJ1734+6052 has two distinct absorption lines at each of the two NaD doublet transitions, as well as at CaK.  Gravitational redshifts are from the Montreal White Dwarf Database \citep[][\url https://www.montrealwhitedwarfdatabase.org/evolution.html]{dufourMWDD} calculated with atmospheric parameters from Table \ref{parameters}. Typical EW uncertainties are $\sim$ 10-20\%.  The typical uncertainty for gravitational redshift is 3 km s$^{-1}$ as realized by propagating from upper and lower bounds of the Teff/log$g$ uncertainties from Table \ref{parameters}. Systemic RV uncertainties can be calculated as an additive combination (in quadrature) of the 3 km s$^{-1}$ from gravitational redshift with the standard deviation of photospheric lines.   } 
\end{table*}

\clearpage

\startlongtable
\begin{deluxetable*}{lccccccccc}
\centerwidetable
\tablecaption{Photospheric Absorption Line Measurements \label{EWs}}
\tablehead{
\colhead{WD} & \colhead{} & \colhead{0218+3625} & \colhead{22211$-$2525} & \colhead{1244+498} & \colhead{1248+1005} & 
\colhead{1922+4709} & \colhead{1734+6052} & \colhead{1415+234} & \colhead{2248+2632} \\
\colhead{T$_{\rm eff}$} & \colhead{} & \colhead{14700 K} & \colhead{14740 K} & \colhead{15150 K} & \colhead{15180 K} &
\colhead{15500 K} & \colhead{16340 K} & \colhead{17300 K} & \colhead{17370 K} \\
\hline
\colhead{Ion} & \colhead{$\lambda$ ({\AA})} & \multicolumn{8}{c}{Equivalent Width (m\AA)}   }
\startdata																													
\ion{O}{1}	&	7771.944	&	200	(27)	&	134	(15)		&	138	(22)	&	266	(59)	&	152	(13)		&	93	(22)	&	80	(11)	&	43	(18)	\\
\ion{O}{1} 	&	7774.166	&	129	(30)	&	88	(16)		&	109	(26)	&	106	(28)	&	124	(15)		&	31	(27)	&	35	(9)	&	31	(13)	\\
\ion{O}{1}	&	7775.388	&	100	(19)	&	81	(19)		&	55	(33)	&	71	(21)	&	82	(12)		&	19	(7)	&	29	(10)	&			\\
\ion{O}{1} 	&	8446.359	&	149	(57)	&	167	(33)		&	75	(22)	&	395	(141)	&	134	(26)		&			&	63	(34)	&			\\
\ion{Na}{1}	&	5889.951	&	45	(10)	&				&			&			&				&			&			&			\\
\ion{Mg}{1}	&	3829.355	&			&	11	(3)		&			&			&				&			&			&			\\
\ion{Mg}{1}	&	3832.304	&	31	(5)	&	42	(3)		&	15	(5)	&	43	(8)	&	39	(6)		&			&	11	(3)	&			\\
\ion{Mg}{1}	&	3838.292	&	58	(4)	&	74	(3)		&	65	(8)	&	76	(16)	&	73	(7)		&			&	47	(7)	&			\\
\ion{Mg}{1}	&	5172.684	&			&	9	(2)		&			&			&	15	(4)		&			&			&			\\
\ion{Mg}{1}	&	5183.604	&	17	(3)	&	31	(4)		&			&			&	32	(10)		&			&			&			\\
\ion{Mg}{2}	&	4481\tablenotemark{$\dagger$}	&	321	(11)	&	400	(8)		&	216	(11)	&	368	(42)	&	374	(25)		&	160	(15)	&	276	(8)	&	97	(13)	\\
\ion{Mg}{2}	&	7877.054	&			&	208	(64)		&			&	179	(68)	&	366	(49)		&			&	91	(41)	&			\\
\ion{Mg}{2}	&	7896.366	&	309	(72)	&	337	(51)		&			&	230	(82)	&	473	(40)		&			&	195	(41)	&	104	(36)	\\
\ion{Al}{2}	&	3587\tablenotemark{$\dagger$}	&	163	(25)	&	39	(14)		&			&			&	241	(45)		&			&			&			\\
\ion{Si}{2}	&	3853.665	&	19	(4)	&	14	(3)		&			&			&	35	(6)		&	8	(4)	&	11	(4)	&			\\
\ion{Si}{2}	&	3856.018	&	105	(3)	&	83	(5)		&	54	(6)	&	72	(5)	&	186	(6)		&	33	(5)	&	82	(3)	&	30	(6)	\\
\ion{Si}{2}	&	3862.595	&	70	(2)	&	48	(4)		&	19	(4)	&	46	(5)	&	128	(8)		&	21	(4)	&	58	(3)	&	20	(3)	\\
\ion{Si}{2}	&	4128.054	&	59	(3)	&	36	(5)		&	13	(5)	&	25	(7)	&	133	(13)		&	13	(4)	&	39	(5)	&			\\
\ion{Si}{2}	&	4130.894	&	92	(3)	&	56	(4)		&	35	(5)	&	80	(13)	&	192	(12)		&	21	(4)	&	58	(5)	&			\\
\ion{Si}{2}	&	5041.024	&	41	(8)	&	28	(9)		&			&			&	152	(13)		&			&	36	(7)	&			\\
\ion{Si}{2}	&	5055.984	&	85	(6)	&	58	(12)		&	30	(8)	&	114	(17)	&	203	(27)		&	27	(8)	&	54	(8)	&			\\
\ion{Si}{2}	&	5957.559	&			&				&			&			&	27	(7)		&			&			&			\\
\ion{Si}{2}	&	5978.930	&			&				&			&			&	81	(13)		&			&			&			\\
\ion{Si}{2}	&	6347.109	&	232	(14)	&	200	(16)		&	104	(10)	&	217	(37)	&	413	(13)		&	126	(37)	&	186	(12)	&	97	(7)	\\
\ion{Si}{2}	&	6371.371	&	153	(12)	&	103	(12)		&	46	(7)	&	158	(29)	&	255	(10)		&	63	(7)	&	89	(11)	&	50	(6)	\\
\ion{Ca}{2}	&	3158.869	&	158	(7)	&	136	(8)		&	128	(8)	&	255	(12)	&	187	(11)		&	54	(7)	&	77	(7)	&	26	(5)	\\
\ion{Ca}{2}	&	3179.331	&	175	(15)	&	181	(10)		&	154	(13)	&	379	(15)	&	246	(22)		&	82	(15)	&	93	(8)	&	47	(6)	\\
\ion{Ca}{2}	&	3181.275	&	52	(6)	&	43	(9)		&			&	41	(7)	&	52	(13)		&	6	(3)	&			&			\\
\ion{Ca}{2}	&	3706.024	&	39	(12)	&	31	(4)		&			&	87	(8)	&	66	(8)		&			&	9	(2)	&			\\
\ion{Ca}{2}	&	3736.902	&	86	(3)	&	86	(3)		&	93	(6)	&	160	(6)	&	137	(10)		&	19	(4)	&	27	(3)	&			\\
\ion{Ca}{2}	&	3933.663	&	595	(20)	&	710	(11)		&	664	(33)	&	1245	(38)	&	528	(23)		&	256	(8)	&	274	(4)	&	169	(7)	\\
\ion{Ca}{2}	&	3968.469	&	338	(15)	&	391	(13)		&	430	(59)	&	747	(50)	&	284	(17)		&	149	(3)	&	154	(4)	&	111	(4)	\\
\ion{Ca}{2}	&	8498.023	&			&				&			&	173	(53)	&				&			&			&			\\
\ion{Ca}{2}	&	8542.091	&	305	(61)	&	305	(40)		&	228	(24)	&	528	(57)	&	255	(18)		&	120	(17)	&	60	(26)	&	88	(31)	\\
\ion{Ca}{2}	&	8662.141	&	192	(50)	&	193	(27)		&	175	(24)	&	386	(52)	&	96	(16)		&	53	(16)	&	45	(12)	&			\\
\ion{Ti}{2}	&	3168.518	&			&				&			&			&				&			&			&			\\
\ion{Ti}{2}	&	3234.520	&	20	(2)	&	15	(2)		&			&	45	(8)	&				&			&			&			\\
\ion{Ti}{2}	&	3236.578	&	16	(3)	&	10	(2)		&			&	40	(7)	&				&			&			&			\\
\ion{Ti}{2}	&	3239.044	&			&	11	(3)		&			&	34	(7)	&				&			&			&			\\
\ion{Ti}{2}	&	3241.994	&	9	(2)	&	9	(2)		&			&	18	(4)	&				&			&			&			\\
\ion{Ti}{2}	&	3248.598	&			&				&			&	16	(7)	&				&			&			&			\\
\ion{Ti}{2}	&	3322.941	&			&				&			&	35	(7)	&				&			&			&			\\
\ion{Ti}{2}	&	3341.880	&	16	(3)	&	8	(3)		&			&	36	(4)	&				&			&			&			\\
\ion{Ti}{2}	&	3349.037	&	13	(2)	&	8	(2)		&	14	(9)	&	47	(11)	&	26	(5)		&			&			&			\\
\ion{Ti}{2}	&	3349.408	&	36	(3)	&	26	(2)		&	19	(4)	&	56	(6)	&	48	(7)		&			&			&			\\
\ion{Ti}{2}	&	3361.218	&	17	(3)	&	13	(2)		&			&	48	(4)	&				&			&			&			\\
\ion{Ti}{2}	&	3372.800	&	16	(22)	&	11	(2)		&			&	30	(3)	&				&			&			&			\\
\ion{Ti}{2}	&	3383.768	&	12	(4)	&	11	(2)		&			&	29	(4)	&				&			&			&			\\
\ion{Ti}{2}	&	3387.846	&			&				&			&	34	(13)	&				&			&			&			\\
\ion{Ti}{2}	&	3685.189	&	17	(6)	&	14	(4)		&			&	29	(5)	&				&			&			&			\\
\ion{Ti}{2}	&	3759.296	&	11	(2)	&	6.3	(1.6)		&			&	16	(3)	&				&			&			&			\\
\ion{Ti}{2}	&	3761.323	&	6	(1)	&	6.1	(1.3)		&			&	13	(3)	&				&			&			&			\\
\ion{Cr}{2}	&	3118.646	&	13	(4)	&	18	(4)		&			&	31	(8)	&	19	(6)		&			&	12	(3)	&			\\
\ion{Cr}{2}	&	3120.359	&	28	(5)	&	30	(4)		&			&	31	(7)	&	38	(9)		&			&	12	(2)	&			\\
\ion{Cr}{2}	&	3124.973	&	32	(5)	&	37	(4)		&	27	(6)	&	43	(9)	&	40	(11)		&			&	27	(6)	&			\\
\ion{Cr}{2}	&	3132.053	&	40	(5)	&	44	(4)		&	28	(5)	&	53	(7)	&	65	(9)		&			&	32	(3)	&			\\
\ion{Cr}{2}	&	3147.220	&			&	9	(3)		&			&	14	(4)	&				&			&			&			\\
\ion{Cr}{2}	&	3180.693	&	18	(4)	&	16	(4)		&			&			&				&			&			&			\\
\ion{Cr}{2}	&	3197.075	&	8	(3)	&	8	(3)		&			&			&				&			&			&			\\
\ion{Cr}{2}	&	3368.041	&	19	(2)	&	15	(2)		&			&	22	(3)	&	28	(7)		&			&	12	(3)	&			\\
\ion{Cr}{2}	&	3408.757	&	15	(3)	&	13	(2)		&			&			&				&			&	9	(2)	&			\\
\ion{Cr}{2}	&	3422.732	&	13	(2)	&				&			&			&				&			&			&			\\
\ion{Cr}{2}	&	3433.295	&	9	(2)	&				&			&			&				&			&			&			\\
\ion{Mn}{2}	&	3441.988	&	17	(2)	&				&			&		    &			    &			&			&			\\
\ion{Mn}{2}	&	3460.316	&	15	(3)	&				&			&			&				&			&			&			\\
\ion{Fe}{1}	&	3570.097	&			&				&			&			&	25	(7)		&			&			&			\\
\ion{Fe}{1}	&	3581.195	&	7	(2)	&				&			&			&	34	(5)		&			&			&			\\
\ion{Fe}{1}	&	3734.864	&	5	(2)	&				&			&			&	25	(5)		&			&			&			\\
\ion{Fe}{1}	&	3749.485	&			&				&			&			&	17	(5)		&			&			&			\\
\ion{Fe}{2}	&	3135.360	&    	&	25	(4)		&			&			&	54	(9)		&			&	16	(3)	&			\\
\ion{Fe}{2}	&	3144.752	&			&	10	(3)		&			&			&	23	(8)		&			&			&			\\
\ion{Fe}{2}	&	3154.202	&	44	(4)	&	61	(5)		&	67	(17)	&	62	(10)	&	79	(10)		&	9	(4)	&	40	(3)	&			\\
\ion{Fe}{2}	&	3162.798	&			&	11	(3)		&			&			&	40	(6)		&			&			&			\\
\ion{Fe}{2}	&	3167.857	&	36	(7)	&	23	(2)		&	37	(14)	&	20	(4)	&	77	(9)		&			&	32	(4)	&			\\
\ion{Fe}{2}	&	3170.337	&			&				&			&			&	24	(6)		&			&			&			\\
\ion{Fe}{2}	&	3177.532	&	22	(6)	&	17	(3)		&	17	(4)	&			&	56	(8)		&			&	34	(6)	&			\\
\ion{Fe}{2}	&	3180.149	&			&				&			&			&	27	(6)		&			&			&			\\
\ion{Fe}{2}	&	3183.111	&	15	(4)	&	16	(5)		&			&			&	38	(7)		&			&	10	(3)	&			\\
\ion{Fe}{2}	&	3186.737	&	16	(4)	&	21	(5)		&	23	(5)	&			&	44	(9)		&			&	14	(4)	&			\\
\ion{Fe}{2}	&	3192.909	&	19	(4)	&	15	(3)		&	12	(3)	&			&	45	(9)		&			&	12	(2)	&			\\
\ion{Fe}{2}	&	3193.799	&	42	(6)	&	32	(4)		&	35	(5)	&	44	(8)	&	79	(7)		&			&	34	(4)	&			\\
\ion{Fe}{2}	&	3196.070	&	18	(3)	&	24	(3)		&	17	(3)	&	27	(6)	&	62	(15)		&			&	14	(3)	&			\\
\ion{Fe}{2}	&	3210.444	&	28	(4)	&	39	(4)		&	29	(5)	&	38	(6)	&	78	(9)		&			&	27	(3)	&			\\
\ion{Fe}{2}	&	3212.017	&			&	8	(3)		&			&			&	41	(8)		&			&			&			\\
\ion{Fe}{2}	&	3213.309	&	55	(4)	&	68	(3)		&	52	(4)	&	41	(4)	&	104	(15)		&	17	(4)	&	39	(3)	&	9	(3)	\\
\ion{Fe}{2}	&	3227.742	&	69	(6)	&	84	(4)		&	97	(5)	&	95	(9)	&	151	(7)		&	26	(5)	&	60	(7)	&	16	(5)	\\
\ion{Fe}{2}	&	3231.706	&			&				&			&			&	24	(5)		&			&			&			\\
\ion{Fe}{2}	&	3232.785	&	8	(2)	&	8	(2)		&	17	(4)	&			&	20	(6)		&			&	10	(3)	&			\\
\ion{Fe}{2}	&	3237.399	&			&				&			&			&	30	(9)		&			&			&			\\
\ion{Fe}{2}	&	3237.820	&	10	(3)	&	7	(2)		&			&	16	(6)	&	32	(7)		&			&	7	(2)	&			\\
\ion{Fe}{2}	&	3243.723	&	11	(5)	&	12	(2)		&	16	(4)	&			&	35	(6)		&			&	9	(3)	&			\\
\ion{Fe}{2}	&	3247.175	&	27	(4)	&	15	(3)		&	28	(6)	&	17	(4)	&	75	(8)		&			&	18	(2)	&			\\
\ion{Fe}{2}	&	3255.887	&	10	(2)	&	11	(2)		&			&			&	37	(6)		&			&	13	(2)	&			\\
\ion{Fe}{2}	&	3258.771	&	12	(2)	&	20	(2)		&	10	(3)	&	18	(5)	&	45	(6)		&			&	21	(4)	&			\\
\ion{Fe}{2}	&	3259.051	&	24	(3)	&	19	(2)		&	20	(4)	&	22	(5)	&	68	(8)		&			&	24	(5)	&			\\
\ion{Fe}{2}	&	3276.604	&			&				&			&			&	20	(5)		&			&			&			\\
\ion{Fe}{2}	&	3277.348	&	13	(2)	&	13	(2)		&			&			&	30	(5)		&			&	9	(2)	&			\\
\ion{Fe}{2}	&	3281.292	&			&				&			&			&	19	(5)		&			&			&			\\
\ion{Fe}{2}	&	3289.354	&			&				&			&			&	28	(5)		&			&			&			\\
\ion{Fe}{2}	&	3323.063	&			&	9	(3)		&			&			&	36	(8)		&			&			&			\\
\ion{Fe}{2}	&	3468.678	&			&				&			&			&	30	(4)		&			&			&			\\
\ion{Fe}{2}	&	3493.470	&	11	(3)	&	12	(2)		&			&			&	37	(6)		&			&	14	(2)	&			\\
\ion{Fe}{2}	&	3748.483	&			&				&			&			&	24	(6)		&			&			&			\\
\ion{Fe}{2}	&	4233.170	&	12	(2)	&	11	(3)		&			&			&	42	(5)		&			&	10	(3)	&			\\
\ion{Fe}{2}	&	4351.769	&			&				&			&			&	23	(6)		&			&			&			\\
\ion{Fe}{2}	&	4522.634	&			&				&			&			&	19	(5)		&			&			&			\\
\ion{Fe}{2}	&	4549.474	&			&				&			&			&	54	(11)		&			&			&			\\
\ion{Fe}{2}	&	4583.837	&	13	(2)	&				&			&			&	50	(7)		&			&			&			\\
\ion{Fe}{2}	&	4923.927	&	23	(10)	&	47	(16)		&	29	(13)	&	32	(11)	&	40	(11)		&	9	(3)	&	13	(3)	&			\\
\ion{Fe}{2}	&	5001.959	&			&				&			&			&	32	(10)		&			&			&			\\
\ion{Fe}{2}	&	5018.440	&	27	(5)	&	31	(11)		&	34	(7)	&	37	(8)	&	68	(10)		&	11	(4)	&	31	(6)	&			\\
\ion{Fe}{2}	&	5035.708	&			&				&			&			&	16	(5)		&			&			&			\\
\ion{Fe}{2}	&	5100.727	&			&				&			&			&	40	(6)		&			&			&			\\
\ion{Fe}{2}	&	5169.033	&	45	(3)	&	56	(6)		&	53	(7)	&	47	(7)	&	143	(12)		&	13	(4)	&	47	(5)	&	8	(2)	\\
\ion{Fe}{2}	&	5197.577	&			&				&			&			&	19	(4)		&			&			&			\\
\ion{Fe}{2}	&	5216.863	&			&				&			&			&	21	(7)		&			&			&			\\
\ion{Fe}{2}	&	5227.481	&			&				&			&			&	78	(10)		&			&			&			\\
\ion{Fe}{2}	&	5234.625	&			&				&			&			&	24	(3)		&			&			&			\\
\ion{Fe}{2}	&	5247.952	&			&				&			&			&	19	(6)		&			&			&			\\
\ion{Fe}{2}	&	5251.233	&			&				&			&			&	21	(7)		&			&			&			\\
\ion{Fe}{2}	&	5260.259	&			&				&			&			&	86	(10)		&			&			&			\\
\ion{Fe}{2}	&	5276.002	&			&				&			&			&	23	(4)		&			&			&			\\
\ion{Fe}{2}	&	5291.666	&			&				&			&			&	21	(5)		&			&			&			\\
\ion{Fe}{2}	&	5316.615	&	15	(3)	&				&	18	(7)	&			&	51	(6)		&			&			&			\\
\ion{Fe}{2}	&	5339.585	&			&				&			&			&	47	(16)		&			&			&			\\
\ion{Fe}{2}	&	5362.869	&			&				&			&			&	17	(3)		&			&			&			\\
\ion{Fe}{2}	&	5506.195	&			&				&			&			&	45	(11)		&			&			&			\\
\enddata
\tablecomments{Wavelengths are in air. EW measurements and uncertainty estimates were made using IRAF's task {\it splot} as described in \citet{klein2021}.}
\tablenotetext{\dagger}{blended multiplet $-$ the EW is the total for the blended feature.}
\end{deluxetable*}

\clearpage

\begin{figure}
\begin{center}
\includegraphics[width=7.5 in]{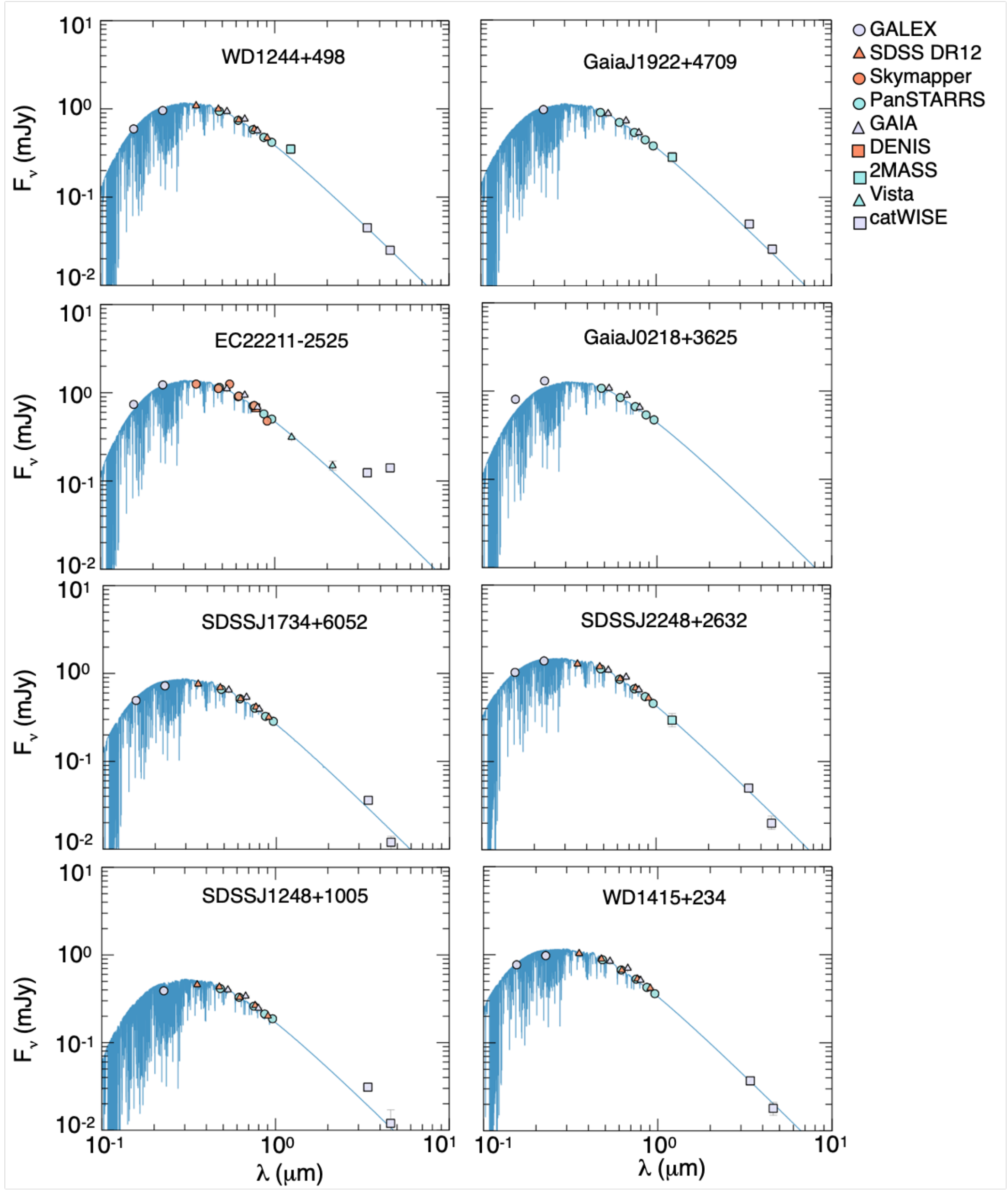}
\caption{SEDs for the eight DB WDs in this study.}
\label{SEDs}
\end{center}
\end{figure}

\renewcommand{\thetable}{A\arabic{table}}
\begin{table*}
\caption{Diffusion timescales and accretion rates for WDs in this study \label{taus}}
\hspace*{-3.7cm}
\begin{tabular}{lccccccccccc}
\hline
\hline
Name   & & $\tau_{\rm Al}$ & $\tau_{\rm Ca}$ & $\tau_{\rm Mg}$ & $\tau_{\rm Si}$ & $\tau_{\rm Fe}$ & $\tau_{\rm O}$ & $\tau_{\rm Na}$& $\tau_{\rm Ti}$& $\tau_{\rm Cr}$& $\tau_{\rm Mn}$ \\
&&Myr&&&&&&&&& \\
\hline
GaiaJ0218$+$3625 && 1.71 & 1.19 & 1.77 & 1.73 & 1.16 & 1.77 & 1.70&	1.09&	1.12&	1.12	\\
WD1244$+$498 && 0.86 & 0.60 & 0.90 & 0.87 & 0.59 & 0.90&&0.55&	0.57&		\\
SDSSJ1248$+$1005 && 0.47 & 0.32 & 0.49 & 0.47 & 0.31 & 0.48&&0.29&	0.30&	0.30	\\
WD1415$+$234 && 0.11 & 0.09 & 0.13 & 0.11 & 0.08 & 0.13&&&	0.08&		\\
SDSSJ1734$+$6052 && 0.35 & 0.25 & 0.37 & 0.33 & 0.24 & 0.38&&&	&		\\
GaiaJ1922$+$4709 && 0.77 & 0.53 & 0.81 & 0.76 & 0.53 & 0.81 & &0.50&	0.51&	0.51	\\
EC22211$-$2525 && 1.42 & 0.98 & 1.47 & 1.43 & 0.96 & 1.47&& &	&		\\
SDSSJ2248$+$2632 && 0.19 & 0.15 & 0.21 & 0.18 & 0.14 & 0.23&&&	&		\\
\hline
Name &${M}_{\rm CV}$& $\dot{M}_{\rm Al}$ &$\dot{M}_{\rm Ca}$ & $\dot{M}_{\rm Mg}$ & $\dot{M}_{\rm Si}$ & $\dot{M}_{\rm Fe}$ & $\dot{M}_{\rm O}$& $\dot{M}_{\rm Na}$& $\dot{M}_{\rm Ti}$& $\dot{M}_{\rm Cr}$& $\dot{M}_{\rm Mn}$  \\
&g&g s$^{-1}$&&&&&&&&&\\
\hline
GaiaJ0218$+$3625 &4.4$\times$10$^{27}$& 2.77$\times$10$^{7}$&1.83$\times$10$^7$ & 1.10$\times$10$^8$ & 1.79$\times$10$^8$ & 2.40$\times$10$^8$ & 9.34$\times$10$^8$&3.72$\times$10$^7$&	5.77$\times$10$^5$&	3.39$\times$10$^6$&	2.50$\times$10$^6$	\\
WD1244$+$498 &2.3$\times$10$^{27}$&5.86$\times$10$^{7^a}$&8.15$\times$10$^7$ & 3.42$\times$10$^8$ & 3.42$\times$10$^8$ & 1.98$\times$10$^9$ & 2.22$\times$10$^9$&&2.89$\times$10$^6$&	1.12$\times$10$^7$ & \\
SDSSJ1248$+$1005 &1.4$\times$10$^{27}$&2.21$\times$10$^{8^a}$&3.14$\times$10$^8$ & 8.46$\times$10$^8$ & 6.60$\times$10$^8$ & 1.79$\times$10$^9$ & 4.89$\times$10$^9$&&1.03$\times$10$^7$&	2.71$\times$10$^7$&	2.56$\times$10$^6$ \\
WD1415$+$234 &2.6$\times$10$^{26}$&1.68$\times$10$^{7^a}$&2.35$\times$10$^7$ & 3.83$\times$10$^8$ & 1.93$\times$10$^8$ & 1.15$\times$10$^9$ & 3.97$\times$10$^8$&&&1.24$\times$10$^7$& \\
SDSSJ1734$+$6052 &3.7$\times$10$^{26}$&2.23$\times$10$^{6^a}$&2.89$\times$10$^6$ & 1.90$\times$10$^7$ & 1.12$\times$10$^7$ & 3.89$\times$10$^7$ & 6.07$\times$10$^7$&&&& \\
GaiaJ1922$+$4709 &1.9$\times$10$^{27}$& 7.01$\times$10$^7$ & 3.31$\times$10$^7$ & 3.29$\times$10$^8$ & 5.29$\times$10$^8$ & 2.13$\times$10$^9$ & 9.14$\times$10$^8$&&1.29$\times$10$^6$&	7.65$\times$10$^6$&	3.98$\times$10$^6$\\
EC22211$-$2525 &3.8$\times$10$^{27}$&1.25$\times$10$^7$& 1.73$\times$10$^7$ & 1.51$\times$10$^8$ & 1.27$\times$10$^8$ & 2.53$\times$10$^8$ & 5.74$\times$10$^8$&&4.07$\times$10$^5$&2.73$\times$10$^6$&1.21$\times$10$^6$ \\
SDSSJ2248$+$2632 &3.8$\times$10$^{26}$&9.05$\times$10$^{6^a}$&1.34$\times$10$^7$ & 4.50$\times$10$^7$ & 2.66$\times$10$^7$ & 4.22$\times$10$^7$ & 1.14$\times$10$^8$&&&& \\
\hline
\multicolumn{12}{l}{Diffusion timescales are from the Montreal White Dwarf Database \citep[MWDD;][]{dufourMWDD}. Mass fluxes are calculated as described}\\
\multicolumn{12}{l}{ in Section \ref{parentbody} for the steady-state accretion phase. ${M}_{\rm CV}$ is the convection zone mass.}\\
\multicolumn{12}{l}{$^a$ Al abundance approximated from a chondritic Al/Ca ratio.}
\end{tabular}
\end{table*}

\clearpage

\bibliographystyle{aasjournal}
\bibliography{bibliography.bib}

\end{document}